\documentclass[a4paper, amsfonts, amssymb, amsmath, reprint, prapplied]{revtex4-2}

\usepackage[english]{babel}
\usepackage[utf8]{inputenc}
\usepackage[T1]{fontenc}
\usepackage{graphicx,import}
\usepackage{units}
\usepackage{layouts}

\usepackage[pdftex, pdftitle={Article}, pdfauthor={Author}]{hyperref} % For hyperlinks in the PDF
\renewcommand{\rm}[1]{\mathrm{#1}}

\renewcommand{\u}[1]{\,\mathrm{#1}}

\usepackage{xcolor}

\usepackage[capitalize]{cleveref}

\DeclareMathOperator*{\argmin}{arg\,min}
\newcommand{\norm}[1]{\left\lVert#1\right\rVert}

\usepackage{tikz}
\usetikzlibrary{matrix}
\usetikzlibrary{fit}
\usetikzlibrary{backgrounds}
\usetikzlibrary{positioning}
\usetikzlibrary{arrows}
\usetikzlibrary{shapes}

\begin{document}

\title{Data-Driven Forecasting of Non-Equilibrium Solid-State Dynamics}

\author{Stefan Meinecke}
    \email[Correspondence email address: ]{meinecke@tu-berlin.de}
    \affiliation{Institut für Theoretische Physik, Technische Universität Berlin, Hardenbergstr. 36, 10623 Berlin, Germany}

\author{Felix Köster}
    \affiliation{Institut für Theoretische Physik, Technische Universität Berlin, Hardenbergstr. 36, 10623 Berlin, Germany}

\author{Dominik Christiansen}
    \affiliation{Institut für Theoretische Physik, Technische Universität Berlin, Hardenbergstr. 36, 10623 Berlin, Germany}

\author{Kathy Lüdge}
    \email[Correspondence email address: ]{kathy.luedge@tu-ilmenau.de}
    \affiliation{Technische Universität Ilmenau, Institut für Physik, Weimarer Straße 25, 98693 Ilmenau, Germany}

\author{Andreas Knorr}
    \affiliation{Institut für Theoretische Physik, Technische Universität Berlin, Hardenbergstr. 36, 10623 Berlin, Germany}

\author{Malte Selig}
    \affiliation{Institut für Theoretische Physik, Technische Universität Berlin, Hardenbergstr. 36, 10623 Berlin, Germany}

\begin{abstract}

We present a data-driven approach to efficiently approximate nonlinear transient dynamics in solid-state systems. 
Our proposed machine-learning model combines a dimensionality reduction stage with a nonlinear vector autoregression scheme. We report an outstanding time-series forecasting performance combined with an easy to deploy model and an inexpensive training routine.
Our results are of great relevance as they have the potential to massively accelerate multi-physics simulation software and thereby guide to future development of solid-state based technologies.

\end{abstract}


\maketitle

\section{Introduction}

Demanding and complex solid-state dynamics are at the core of many challenging multi-physics problems. Examples include solar cells \cite{LIU02b}, semiconductor lasers \cite{LIN13,MUN17,KAN17a,KAN20a,HAU21,THU21}, mode-locked semiconductor lasers \cite{ROS11c,KIL14,MEI19,MCL20}, semiconductor optical amplifiers \cite{KOL13,CAP13,LIN18}, and nano-photonic devices based an plasmonic nanostructures \cite{franke2019quantization,carlson2021strong}.
Often, the dynamics of excited charge carriers and phonons themselves are not at the center of attention but, nonetheless, need to be solved by multi-physics simulation codes, e.g., to acquire the needed degree of accuracy to calculate macroscopic dynamics. In order to maintain reasonable computational costs, microscopic solid state dynamics are then either completely eliminated or treated within some kind of approximation, e.g., the relaxation-time approximation\cite{Czycholl}. Such approaches come at the cost of predictive power, but have -  nevertheless - revealed countless valuable insights into multi-physics phenomena (see previous references for examples).

In this manuscript, we intend to fill the gap between computationally expensive microscopic calculations and rough analytic approximations by proposing a numerically efficient data-driven approximation scheme. Our approach is highly motivated by the immensely successful application of the machine-learning paradigm to tasks in science, technology, and everyday life \cite{bertolini2021machine,sarker2021machine,zhang2019reference,wang2021deep,lee2021short,de2021simulation,zhang2019deep}.
Within science, data-driven approaches have been especially popular among the data-rich fields such as
particle physics\cite{radovic2018machine}, fluid dynamics\cite{brunton2020machine}, astrophysics\cite{ourmazed2020}, and X-ray free electron laser experiments\cite{chen2021machine,xian2020machine}. 
Similarly, other scientific fields such as biology\cite{baldi2001bioinformatics}, medicine\cite{may2021eight}, and chemistry\cite{noordik2004cheminformatics} have been deploying data-driven methods for some time. 
However, data-driven approaches only start to slowly rise in quantum optics\cite{palmieri2020experimental,kaestle2021sampling} and solid-state material science \cite{schmidt2019recent}. A few examples include the prediction of static quantities to predict new stable materials\cite{lookman2018materials,ryan2018crystal,graser2018machine,balachandran2018experimental,oliynyk2016classifying,li2018predicting,ward2017including,faber2016machine}, the prediction of material parameters\cite{xie2018crystal,ju2017designing,seko2017representation}, and speed-up of first principle calculations\cite{jalem2018bayesian}. 

For our purpose of developing a data-driven approximation for solid-state dynamics, we devise a two-dimensional coupled electron-phonon system as a toy model. For convenience, we consider parameters for transition metal dichalcogenide monolayers\cite{Li2013,Jin2014,Steinhoff2014,perea2019exciton,selig2019ultrafast}, MoSe$_2$ in our case.
In thermal equilibrium, the electron system follows a Fermi-Dirac distribution and the phonon system a Bose-Einstein distribution, both with a common temperature $T$. After such a system has been subject to a perturbation, it exhibits transient dynamics during its thermalization. For weak perturbations, analytic approximations - the relaxation time approximation which assigns a single number to the complex relaxation process -  have been successfully developed\cite{Czycholl}. Strong perturbations, e.g., as they are induced by powerful optical pulses, however, defy analytic efforts so far and thus represent this works target.
We therefore consider the task of forecasting the transient dynamics of the electron system from a given strongly perturbed initial state. 
The focus is given to the electrons, since they easily couple to another physical system, e.g., via a macroscopic electric field. The calculated phonon dynamics, on the other hand, is intentionally hidden from the data-driven model.
Note that this renders the observed electron dynamics non-Markovian, since the hidden phonon system augments the electron trajectory with a path dependence.
Our investigated electron-phonon system can therefore represent a wide class of problems: The electron system represents the subsystem, which couples to another physical system, and is therefore of primary interest. The phonon system, on the other hand, is only of secondary interest, since it represents some internal dynamics, which affect the interaction with the other physical system only indirectly via the coupling subsystem (the elecrons in our case).

With the goal of accelerating the simulation of a multi-physics problem by replacing the expensive computation of the complex nonlinear system (the coupled electron-phonon system) by a data-driven approximation, we put a strong constraint on the latter. This approach becomes only viable, if the combined efforts of generating the training data; implementing, training, and testing the data-driven model; and performing the forecasting are considerably less than the direct simulation in order to justify the potential loss of accuracy. Hence, one must be careful and deliberate with the choice of the forecasting approach from the vast number of options that have been developed in the past decades.

Traditionally, time-series forecasting has almost been synonymous with autoregressive (AR) models and variants thereof such as autoregressive moving-average models (ARMA) or vector autoregressive (VAR) models \cite{BIL13,BOX15}. These models propagate the system based on a linear combination of past system states. By feeding their output back as input, they yield autonomous dynamical systems and can generate forecasts of arbitrary length. Such models, however, perform unsurprisingly poorly for strongly nonlinear systems.
To overcome this issue, one popular approach has been the usage of artificial recurrent neural networks, both in the manifestation of \emph{long short-term memory} networks (LSTMs) \cite{HOC97,GEN16a,ZHA17e,VLA18,CHI20a,VLA20a,WAN21b} and \emph{echo state networks} (ESNs)/\emph{reservoir computers} \cite{JAE01,LUK09,PAT18,CHA20b,VLA20a,ROE21,PYL21}. Both, however, come with their specific limitations: LSTMs, e.g., are comparatively expensive to train, both in terms of training data and computational costs \cite{CHA20b} and reservoir computers require tedious hyper-parameter optimizations and long warm-up times \cite{GAU21b}.
On that account, models based on nonlinear vector autoregression (NVAR) have recently demonstrated excellent forecasting abilities with inexpensive training costs, easy deployability, and negligible warm-up time \cite{PYL21,GAU21b,BOL21,CHE22,BAR22a}. 
For those reasons, we follow this approach and develop a nonlinear autoregressive reduced-order model (NARROM) by complementing the aforementioned NVAR model with dimensionality reduction stage. This allows us to greatly reduce the computational demands by performing the forecasting in a reduced-order latent space.

In this work, we discuss the design, the implementation, the application to the coupled electron-phonon system, and the optimization of the nonlinear autoregressive reduced-order model. Our results demonstrate that a well tuned model can excellently forecast the high-dimensional nonlinear transient dynamics, which are well beyond the scope of analytic approximations. We therefore believe that our data-driven approach to the approximation of non-equilibrium solid-state dynamics has the potential to greatly accelerate the simulation of challenging multi-physics problems.

The manuscript is organized as follows: \Cref{sec:dynamics} introduces and illustrates non-equilibrium transient dynamics of the coupled electron-phonon system. \Cref{sec:model} rigorously walks through the design, training, and operation of the nonlinear autoregressive reduced-order model. The forecasting abilities of the model are benchmarked and analyzed in detail in \cref{sec:results} for two different nonlinear transformations. Lastly, the results are summarized and discussed in \cref{sec:discussion}.

\section{Electron-Phonon Dynamics} \label{sec:dynamics}

To study the coupled electron-phonon dynamics, we numerically solve the Boltzmann scattering equations for electrons and phonons (s. \cref{sec:eom}). Throughout this manuscript, we solve the Boltzmann equations for two-dimensional MoSe$_2$ as an exemplary material, taking into account the $LA$, $TA$, $TO$ and $A'$ phonon branches. The dispersion of the acoustic modes is treated in the Debye approximation with velocities of sound taken from ab initio calculations\cite{Li2013}. The dispersion of of the optical modes is treated in the Einstein approximation with constant energies taken from ab initio calculations\cite{Li2013}. The full parameter set of our evaluation is given in \cref{sec:eom}. We note, that although excitons dominate the properties of transition metal dichalcogenides, we ignore the influence of them on the dynamics, as the scope of our work is to demonstrate the applicability of the data-driven approach to the semiconductor dynamics. As initial conditions for the electrons, we consider Gaussians $f_\mathbf{k}^0 = A \exp(\frac{-\mathbf{k}^2 - 2 m \tilde{E}}{4m\sigma^2})$ with varying height $A$, center energy $\tilde{E}$ and width $\sigma$ of the distribution. Such distributions appear for instance after strong off-resonant excitations\cite{selig2022impact}. The phonons are initialized with a Bose-Einstein distribution at $T= \unit[300]{K}$.

\begin{figure}[htbp]
\centering
\includegraphics[width=\linewidth]{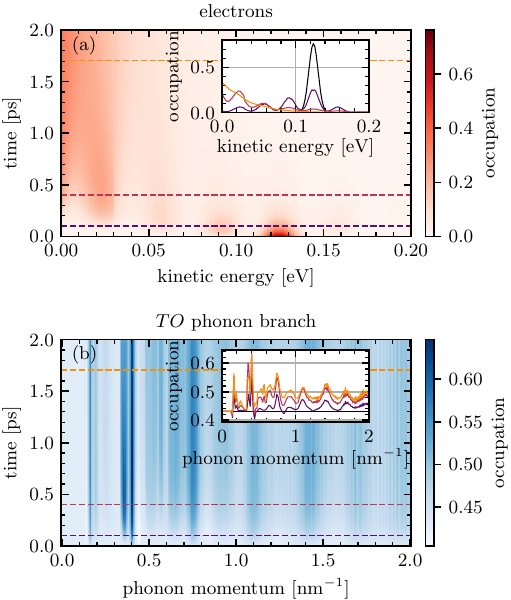}
\caption{Color-coded dynamics of the electron occupations (a) and the $TO$ phonon branch occupation (b). Insets show slices of the respective distributions at the times $t \in \{0,0.1,0.4,1.7\}$\,ps (indicated by dashed horizontal lines).
}
\label{fig:dynamics}
\end{figure}

\Cref{fig:dynamics} illustrates an exemplary time trace of the thermalization of the electron-phonon system. \Cref{fig:dynamics}\,(a) shows the temporal evolution of the electron occupation. The inset illustrates snapshots of the electron occupation at selected times. As electron-phonon scattering sets in, the electrons scatter down to lower energies. Interestingly, in the transient regime before thermalization, fringes in the electron distribution with a period of \unit[30]{meV} can be observed, which originate from the fast relaxation of electrons under the emission of optical phonons. On short time scales, this relaxation channel dominates over the relaxation assisted by acoustic phonons, due to the larger energies of the involved phonons. However, acoustic phonons act on longer timescales leading mainly to a smoothing of the electron distribution and to the relaxation of electrons with kinetic energies smaller than the optical phonon energy. The electrons reach a quasi Fermi-Dirac distribution after $\approx\unit[2]{ps}$. 
\cref{fig:dynamics}\,(b) illustrates the temporal evolution of the phonon occupation of the $TO$ mode. We find that for most momenta, the occupation increases due to phonon emission processes, which indicates the energy transfer from the electron system to the phonon system, i.e., a heating of the phonon system. We find an enhanced formation of optical phonons at certain momenta, which correspond to those momenta, where energy and momentum conservation are simultaneously fulfilled during the electron-phonon scattering (s. \cref{eq:elec} and \cref{eq:phon}).

\section{Nonlinear autoregressive Reduced Order Model} \label{sec:model}

For the efficient forecasting of the electron dynamics at discrete times, we propose and develop a data-driven nonlinear autoregressive reduced order model (NARROM). Our approach combines a dimensionality reduction scheme with a nonlinear vector autoregressive (NVAR) model \cite{LUE05,BOL21,GAU21b}. Hence, it can be broken apart into two independent parts:

Firstly, the dimensionality reduction is achieved by truncating the expansion of the system state in a suitable basis. This is motivated by the fact that all transient dynamics, despite their complexity and diversity, are constraint by the laws of physics, i.e., the electron-phonon interaction and therefore only occupy a small subspace of the electron configuration space. We therefore expect low dimensional patterns in the transient dynamics and thus aim to construct a basis to capture those with only a few relevant modes. Specifically, we want the aforementioned subspace to be well represented by a few dominant modes, such that expansions of transient states quickly converge and an appropriate truncation only produces small errors. This, however, also means that we generally expect this dimensionality reduction approach to be lossy.

Secondly, the nonlinear vector autoregressive (NVAR) model constructs a feature vector via a nonlinear transform of the past system states. The next system state is then inferred via a linear transformation of this feature vector. The model therefore provides both a memory of past system states and a nonlinearity, which are relevant in the considered electron dynamics, by design. 
Taking into account the past system states can also be thought of as a delay embedding in the sense of Taken's embedding theorem \cite{TAK81}. Hence, the neglected internal dynamics, i.e., the phonon system in our case, are implicitly taken into account for a sufficient delay embedding dimension.
Note that in order to build an efficient and well performing model, a suitable feature space must be constructed. This is by no means a trivial task, since the possibilities for nonlinear transforms are vast.

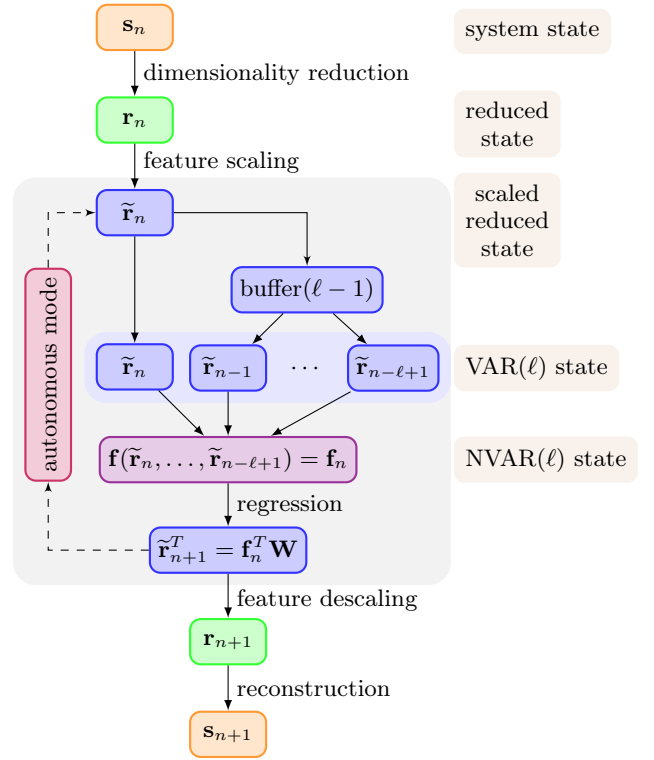
\begin{figure}
\centering
\tikzstyle{state}=[rectangle,
    thick,
    minimum height=0.6cm,
    minimum width=1.0cm,
    draw=orange!80,
    fill=orange!20,
    rounded corners]
    
\tikzstyle{cstate}=[rectangle,
    thick,
    minimum height=0.6cm,
    minimum width=1.0cm,
    draw=green!80,
    fill=green!20,
    rounded corners]
    
\tikzstyle{rcstate}=[rectangle,
    thick,
    minimum height=0.6cm,
    minimum width=1.0cm,
    draw=blue!80,
    fill=blue!20,
    rounded corners]
    
\tikzstyle{dummy}=[rectangle,
    % thick,
    minimum height=0.6cm,
    minimum width=1.0cm,
    % draw=blue!80,
    % fill=blue!20,
    rounded corners]

\tikzstyle{nltcstate}=[rectangle,
    thick,
    minimum height=0.6cm,
    minimum width=1.2cm,
    draw=violet!80,
    fill=violet!20,
    rounded corners]
    
\tikzstyle{auto}=[rectangle,
    thick,
    minimum height=0.6cm,
    minimum width=0.6cm,
    draw=purple!80,
    fill=purple!20,
    rounded corners]
    
\tikzstyle{line} = [draw, -latex']

\tikzstyle{desc}=[rectangle,
    minimum height=0.6cm,
    minimum width=0.6cm,
    fill=brown!10,
    rounded corners]

\tikzstyle{background}=[rectangle,
    fill=gray!10,
    inner sep=0.15cm,
    rounded corners=3mm]
    
\tikzstyle{background_VAR}=[rectangle,
    fill=blue!10,
    inner sep=0.15cm,
    rounded corners=3mm]

\begin{tikzpicture}[>=latex]

  \matrix (mtrx) [row sep=0.6cm, column sep=0.2cm, matrix of nodes, nodes in empty cells] {

    % &  \node (s_n) [state]{$\mathbf{s}_{n}$}; & & & \\
    % &  \node (c_n) [cstate]{$\mathbf{c}_{n}$}; & & & \\
    % &  \node (tc_n) [rcstate]{$\mathbf{\widetilde{c}}_{n}$}; & & & \\
    % & & & \node (tc_n_2) [rcstate]{$\mathbf{\widetilde{c}}_{n}$}; & \node (dots2) {$\cdots$}; \\
    % & & \node (tc_n_31) [rcstate]{$\mathbf{\widetilde{c}}_{n}$}; & \node (tc_n_32) [rcstate]{$\mathbf{\widetilde{c}}_{n-1}$}; & \node (dots) {$\cdots$}; \\
    % & \node (tc_n_41) [rcstate]{$\mathbf{\widetilde{c}}_{n}$}; & \node (tc_n_42) [rcstate]{$\mathbf{\widetilde{c}}_{n-1}$}; & \node (tc_n_43) [rcstate]{$\mathbf{\widetilde{c}}_{n-2}$}; & \node (dots4) {$\cdots$}; \\
    % & & & & \node (d75) [dummy] {}; \\
    % % & &  \node (tc_n+1) [rcstate]{$\mathbf{\widetilde{c}}_{n+1} = \mathbf{Wf}_n$}; \\
    % & & & & \node (d85) [dummy] {};\\
    % & &  \node (c_n+1) [cstate]{$\mathbf{c}_{n+1}$}; \\
    % & & \node (s_n+1) [state]{$\mathbf{s}_{n+1}$}; \\
    % \\
    % };
    
    &  \node (s_n) [state]{$\mathbf{s}_{n}$}; & & & \\
    &  \node (c_n) [cstate]{$\mathbf{r}_{n}$}; & & & \\
    &  \node (tc_n) [rcstate]{$\mathbf{\widetilde{r}}_{n}$}; & & & \\
    & & & & \\

    & \node (tc_n_41) [rcstate]{$\mathbf{\widetilde{r}}_{n}$}; & \node (tc_n_42) [rcstate]{$\mathbf{\widetilde{r}}_{n-1}$};  & \node (dots4) {$\cdots$};  & \node (tc_n_43) [rcstate]{$\mathbf{\widetilde{r}}_{n-\ell+1}$};\\
    & & & & \node (d75) [dummy] {}; \\
    & & & & \node (d85) [dummy] {};\\
    & &  \node (c_n+1) [cstate]{$\mathbf{r}_{n+1}$}; \\
    & & \node (s_n+1) [state]{$\mathbf{s}_{n+1}$}; \\
    \\
    };
    
    % \node (nltcs) [nltcstate, fit=(mtrx-7-2) (mtrx-7-4), text centered, text height = 2mm, text depth = 0mm ]{$\mathbf{f}(\mathbf{\widetilde{c}}_{n}, \mathbf{\widetilde{c}}_{n+1},\dots) = \mathbf{f}_n$};
    
    \node (invpos) [fit=(mtrx-4-1) (mtrx-6-1)]{};
    \node (auto) [auto, left=of invpos,xshift=0.6cm,rotate=90,anchor=north] {autonomous mode};
    
    \draw (mtrx-4-4) node (mem) [rcstate]{buffer$(\ell -1)$};
    
    \draw (mtrx-6-3) node (nltcs) [nltcstate]{$\mathbf{f}(\mathbf{\widetilde{r}}_{n},\dots, \mathbf{\widetilde{r}}_{n-\ell+1}) = \mathbf{f}_n$};
    
    \draw (mtrx-7-3) node (tc_n+1) [rcstate]{$\mathbf{\widetilde{r}}_{n+1}^T = \mathbf{f}_n^T \mathbf{W}$};
    
    \draw (mtrx-1-5) node[desc, right, inner sep=1.5mm, outer sep=8mm] {system state};
    \draw (mtrx-2-5) node[desc, right, inner sep=1.5mm, outer sep=8mm, align=center] {reduced\\state};
    \draw (mtrx-3-5) node[desc, right, inner sep=1.5mm, outer sep=8mm, align=center] {scaled\\reduced\\state};
    \draw (tc_n_43) node[desc, right, inner sep=1.5mm, outer sep=8mm, align=center] {VAR($\ell$) state};
    \draw (d75) node[desc, right, inner sep=1.5mm, outer sep=8mm, align=center] {NVAR($\ell$) state};
    
    \path[->]
        (s_n) edge node[pos=0.5,right] {dimensionality reduction}  (c_n)
        (c_n) edge node[pos=0.4,right] {feature scaling} (tc_n)
        % (tc_n) edge (mem)
        (tc_n) edge (tc_n_41)
        (mem) edge (tc_n_42)
        (mem) edge (tc_n_43)
        
        (tc_n_41) edge (nltcs)
        (tc_n_42) edge (nltcs)
        (tc_n_43) edge (nltcs)
        
        (nltcs) edge node[right] {regression} (tc_n+1)
        (tc_n+1) edge node[pos=0.6,right] {feature descaling} (c_n+1)
        (c_n+1) edge node[right] {reconstruction} (s_n+1)
    ;
    
    % \path [line,dashed] (tc_n_2) --  (tc_n_32);
    \path [line,dashed] (auto) |-  (tc_n);
    \path [line,dashed] (tc_n+1) -|  (auto);
    
    \path [line] (tc_n) -|  (mem);
    
    \begin{pgfonlayer}{background}
        \node [background,
                    fit=(tc_n) (auto) (tc_n_43) (tc_n+1)] {};
        \node [background_VAR,
                    fit=(tc_n_41) (tc_n_43)] {};
    \end{pgfonlayer}
\end{tikzpicture}
\caption{Flow diagram of the \textbf{n}onlinear \textbf{a}uto\textbf{r}egressive \textbf{r}educed \textbf{o}rder \textbf{m}odel (NARROM). $\mathbf{s}_{n}$, $\mathbf{r}_{n}$, and $\mathbf{\widetilde{r}}_{n}$ denote the system, reduced system, and scaled reduced system state vectors at the discrete time $n$. VAR$(\ell)$ state denotes the concatenation of the past $\ell$ state vectors and NVAR state a nonlinear transform $\mathbf{f}(\cdot)$ thereof. The model is trained by optimizing the regression weights $\mathbf{W}$. In autonomous mode, the model output $\mathbf{\widetilde{r}}_{n+1}$ is feed back as an input.}
\label{fig:model}
\end{figure}

In the following, we formalize the proposed model.
The complete structure of the nonlinear autoregressive reduced order model is sketched in \cref{fig:model} as a flow diagram. 
It illustrates the forecasting of the next system state at the time $t_{n+1}$ based on the prior system states.
The current system state at time $t_n$ is represented by the state vector $\mathbf{s}_n \in \mathbb{R}^{d_s}$ (orange box) on the very top. Time itself is assumed to be discretized into equidistant steps $\Delta t$ with $t_n = t_0 + n\Delta t$. 
In the first step, the system state is mapped to the reduced-order vector $\mathbf{r}_n \in \mathbb{R}^{d_r<d_s}$ (green box) via the dimensionality reduction scheme,
Each element of the reduced state vector, i.e., the features of the reduced-order latent state space, is subsequently rescaled and represented by $\mathbf{\widetilde{r}}_n  \in \mathbb{R}^{d_r}$ (blue boxes). This step is crucial, since the magnitudes of the linear features critically determine the response of the following nonlinear transformation. 
Before that, however, the scaled reduced state is also passed to a buffer, which stores the past $(\ell - 1)$ states and provides them for further processing.
The concatenation of the $\ell$ scaled reduced state vectors $\mathbf{\widetilde{r}}_n, \dots, \mathbf{\widetilde{r}}_{n-\ell+1}$ in the blue shaded row then represents the linear VAR$(\ell)$ state \cite{LUE05,QIN11} in the reduced latent space. 
The VAR$(\ell)$ state is passed to a nonlinear function $\mathbf{f}(\cdot)$, which produces the NVAR$(\ell)$ state $\mathbf{f}_n = \mathbf{f}(\mathbf{\widetilde{r}}_{n},\dots,\mathbf{\widetilde{r}}_{n-l+1}) \in \mathbb{R}^{d_f}$ \cite{BIL13} (purple box), i.e., the feature vector of the nonlinear autoregressive reduced order model. 
Next, the regression step is performed via a linear transformation of the feature vector to produce the scaled reduced state vector
\begin{align}
    \mathbf{\widetilde{r}}_{n+1}^T = \mathbf{f}_n^T \mathbf{W} \label{eq:autoregression_step}
\end{align}
at the time $t_{n+1}$. The matrix $\mathbf{W} \in \mathbb{R}^{d_f \times d_r}$ contains the regression weights. Note that transposed vectors (row vectors) are used since the data matrices, on which $\mathbf{W}$ is trained, are constructed from rows of data vectors. 
Lastly, the scaled reduced system state is descaled and the full system state $\mathbf{s}_{n+1}$ is reconstructed in the two most bottom steps of the flow diagram.

The model is trained using the supervised learning paradigm, where the models parameters are optimized to minimize a loss function of some training data and its corresponding model output.
In principle, the training of our proposed model can include the dimensionality reduction stage, the feature scaling stage, and the regression weights $\mathbf{W}$. For the sake of training simplicity, we treat the feature scaling stage as a hyperparameter problem and train the compression stage and the regression weights separately. I.e., we first optimize the dimensionality reduction stage to best reproduce the training data for a given dimensionality $d_r$ and then optimize the regression weights $\mathbf{W}$.

To facilitate the model training, we first sample $M$ trajectories at $N_M$ discrete times and build the data matrix
\begin{align}
    \mathbf{S} = (\mathbf{s}_1^1,\dots,\mathbf{s}_n^m,\dots,\mathbf{s}_{N_M}^M)^T,    
\end{align}
where the rows correspond to the system state vectors $\mathbf{s}_n^m$ of the $m$-th trajectory at the time $t_n^m$.

To optimize the dimensionality reduction stage, we minimize the reconstruction error
\begin{align}
    % \argmin_{\mathbf{B}} \norm{\mathbf{S}\mathbf{B}_\rm{c}\mathbf{B}_\rm{c}^{-1} - \mathbf{S}}_2
    \argmin_{\{R\}} \norm{R^{-1}\left(R\left(\mathbf{S}\right)\right) - \mathbf{S}}_\rm{F}^2,
\end{align}
where $R$ and $R^{-1}$ denote the dimensionality reduction and reconstruction functions and $\norm{\cdot}_\rm{F}$ the Frobenius norm. The optimization parameters $\{R\}$ are specific to the chosen approach and are exemplarily discussed in the \cref{sec:dim_reduction}.

We then proceed to construct the feature matrix
\begin{align}
    \mathbf{F} = (\mathbf{f}_1^1,\dots,\mathbf{f}_n^m,\dots,\mathbf{f}_{N_M-1}^M)^T,     \label{eq:feature_matrix}
\end{align}
where the feature vectors $\mathbf{f}_n^m$ are computed according to \cref{fig:model} from the system states. Each trajectories final state $\mathbf{s}_{N_M}^M$ is omitted and only appears in the target matrix. If necessary ($\ell > 1$), the trajectories $\mathbf{s}_n^m$ are padded with the trajectories initial state $\mathbf{s}_1^m$ to fill up the VAR$(\ell)$ state and calculate the initial feature vectors $\mathbf{f}^m_{n<\ell}$.
Lastly, the target matrix
\begin{align}
    \mathbf{T} = (\mathbf{\widetilde{r}}_2^1,\dots,\mathbf{\widetilde{r}}_n^m,\dots,\mathbf{\widetilde{r}}_{N_M}^M)^T    
\end{align}
is constructed from all scaled reduced system states $\mathbf{\widetilde{r}}_n$ except each trajectories initial state. Hence, the feature matrix and the target matrix are shifted by one time step $\Delta t$ with respect to each other.

The regression weights $\mathbf{W}$ are then determined by solving the least-squares problem
\begin{align}
    \argmin_{\mathbf{W}} \left[ \norm{\mathbf{F}\mathbf{W} - \mathbf{T}}_\rm{F}^2 + \alpha \norm{\mathbf{W}}_\rm{F}^2 \right], \label{eq:lossfunction}
\end{align}
where $\norm{\cdot}_\rm{F}$ is the Frobenius norm and $\alpha$ the Tikhonov regularization (ridge) parameter \cite{VOG02,PRE07,BRU22}. A nonzero $\alpha$ penalizes large weights $w_{nm}$ and thereby counteracts overfitting problems. The solution to this problem is known and reads
\begin{align}
    \mathbf{W} = \left( \mathbf{F}^T \mathbf{F} + \alpha \mathbf{I} \right)^{-1} \mathbf{F}^T \mathbf{T},
\end{align}
where $(\cdot)^{-1}$, depending on the rank of $\left( \mathbf{F}^T \mathbf{F} + \alpha \mathbf{I} \right)$, either denotes the inverse or the Moore–Penrose pseudo inverse.

Once trained, the model can be used to advance the system by one discrete time step $\Delta t$ as shown in \cref{fig:model}. Trajectories of arbitrary length can be produced from the model in autonomous mode, in which the models output is fed back as an input. 
Since the regression error accumulates in this mode, exceptionally good per-step predictions are required to achieve high quality trajectories.

\section{Results} \label{sec:results}

In this section, we construct a complete nonlinear autoregressive reduced order model by specifying the individual stages. 
This includes the dimensionality reduction, the feature scaling, and the nonlinear transformation. We propose multiple approaches for each stage and discuss and compare them in terms of their forecasting accuracy and their computational demands.
For that purpose, we simulate a data set with $n=1000$ individual trajectories, which we use for training and testing. The relevant details can be found in \cref{sec:training_data}.

\subsection{Dimensionality reduction} \label{sec:dim_reduction}

First, we set out to find and optimize a linear dimensionality reduction scheme that works well for our data. As elaborated in \cref{sec:model}, this translates into constructing a basis in which the expansion of the considered system states converges sufficiently with a minimal number of basis vectors. On that account, we evaluate the performance in terms of the reconstruction error for three dimensionality reduction schemes:

In the first case, we use the left singular vectors of the singular value decomposition (SVD) \cite{GOL13a, BRU22}
\begin{align}
    \mathbf{S}^T = \mathbf{U} \mathbf{\Sigma} \mathbf{V}^T, \label{eq:SVD}
\end{align}
which is successfully utilized in many branches of science and engineering \cite{PEA01,HOT33,LUM67,HOL12a,STE93c,BRU22}. The SVD has a long history \cite{GOL13a, BRU22} and is known across different disciplines as the Karhunen–Lo{\`e}ve transform (KLT) \cite{KAR47,LOE17a}, empirical orthogonal functions \cite{LOR56}, proper orthogonal decomposition (POD) \cite{LUM67}, and canonical correlation analysis \cite{CHE96}.

Note that we have transposed our data matrix $\mathbf{S}$ such that the system states are organized as column vectors, in order to conform to the common SVD literature. The matrices $\mathbf{U}$ and $\mathbf{V}$ are unitary and contain the left and right singular vectors of $\mathbf{S}^T$ and $\mathbf{\Sigma}$ is a diagonal matrix with the singular values in descending order. Their magnitude naturally organizes the left and right singular vectors according to their share in reconstructing the data matrix $\mathbf{S}^T$. Specifically, the Eckart-Young theorem \cite{ECK36,BRU22} states that the best rank-$r$ approximation of a matrix with respect to the Frobenius norm can be achieved via the truncation over the leading $r$ singular values of the SVD. 
To reduce a given system state $\mathbf{s}_n$, we therefore compute the SVD of the training data to obtain the basis $\mathbf{U}$. We then use the truncated matrix $\mathbf{U}_{r} = (\mathbf{u}_1,\dots,\mathbf{u}_{r})$
to project $\mathbf{s}_n$ onto the first $r$ left singular vectors $\mathbf{u}_m$ to obtain the reduced state
\begin{align}
    \mathbf{r}_n = (r_1,\dots,r_{r})^T = \mathbf{U}_{r}^T \mathbf{s}_n.
\end{align}
The corresponding reconstructed system state $\hat{\mathbf{s}}_n$ is then obtained via
\begin{align}
    \hat{\mathbf{s}}_n = \mathbf{U}_{r} \mathbf{r}_n =  \mathbf{U}_{r} \mathbf{U}_{r}^T \mathbf{s}_n,
\end{align}
where the hat denotes the expansion of the state in the truncated basis. 
Note, that this dimensionality reduction scheme is entirely data driven. We therefore expect its performance to critically depend on the quality of the training data. This means that the training data should be representative of the relevant system configuration subspace, to generate a basis that is optimally adapted to the task.

In the second case, we expand the system state into the modes of the discrete Fourier transform (DFT). This approach is well established and has the advantages of computational efficiency via the fast Fourier transform (FFT) as well as interpretable modes in terms of their characteristic frequencies. In the case of a real valued system state with dimension $d_s$, we obtain $d_s/2$ unique complex Fourier coefficients. Instead of naively truncating the expansion at some frequency, we split the complex coefficients into their real and imaginary parts and then compute the mean of all $d_s$ coefficients of the transformed training data. This allows us to sort the Fourier modes by the descending order of their mean coefficients and only keep the leading $r$ in the truncation. Note, that this strategy adds a data-driven element to the DFT reduction scheme, in order to better adapt it to the data at hand. The reconstruction is achieved by calculating the inverse fast Fourier transform, where the coefficients of the previously truncated modes are set to zero.

In the last case, we construct a basis using the Hermite-Gauss functions
\begin{align}
    \psi_n(x) =& \left (2^n n! \sqrt{\pi} \right )^{-\frac12} e^{-\frac{x^2}{2}} H_n(x) \nonumber \\
            =& (-1)^n \left (2^n n! \sqrt{\pi} \right)^{-\frac12} e^{\frac{x^2}{2}} \frac{d^n}{dx^n} e^{-x^2},
\end{align}
which form an orthonormal basis of $L^2(\mathbb{R})$. This is motivated by the fact that the zeroth mode
\begin{align}
    \psi_0(x) = \pi^{-\frac14} e^{-\frac{x^2}{2}} 
\end{align} approximates the quasi-equilibrium Fermi-Dirac electron distribution (up to a constant factor), if we substitute $x$ with $\hbar k/\sqrt{2 m k_\rm{B} T}$ and assume large temperatures or low densities. Deviations from the equilibrium state can then be expanded in the higher modes $\psi_n$. To build the desired basis, we sample the Hermite-Gauss functions at the discretization points $k$ to construct the matrix $\widetilde{\mathbf{H}} \in \mathbb{R}^{d_s \times d_s}$. The matrix elements are given by
\begin{align}
    \widetilde{h}_{lm} = \psi_m(l\Delta k),
\end{align}
where $\Delta k$ denotes grid size in $k$-space. The columns $\widetilde{\mathbf{h}}_m$ thus correspond to the different modes. Note that those vectors are in general not orthonormal, unlike the original functions, since they have been obtained via a finite number of samples from the bounded interval $[0,k_\rm{max}]$. We therefore apply a Gram-Schmidt process to $\widetilde{\mathbf{H}}$ to obtain the orthonormal matrix $\overline{\mathbf{H}}  \in \mathbb{R}^{d_s \times d_s}$. This leaves the lower modes almost untouched and mostly modifies the higher modes, which extend to larger $k$ and are thus more affected by the sampling cut-off at $k_\rm{max}$. Similar to the DFT reduction scheme, we reorganize the columns, i.e., the order of the modes, according to the descending order of the transformed training data's expansion coefficients to obtain the matrix $\mathbf{H}$. This way, the first $r$ columns $\mathbf{h}_m$ maximize their share in the expansion of the training data. 
Hence, we can reduce and reconstruct a given system state using the truncated matrix $\mathbf{H}_{r}$, which only contains the first $r$ columns. Lastly, we further optimize this dimensionality reduction scheme by treating the temperature $T$ not as an external parameter, but as an optimization parameter to minimize the Frobenius norm between the reconstructed training data and the training data. The Hermite reduction scheme is thus both physically motivated and subject to data-driven optimizations.

\begin{figure}[htbp]
\centering
\includegraphics[width=\linewidth]{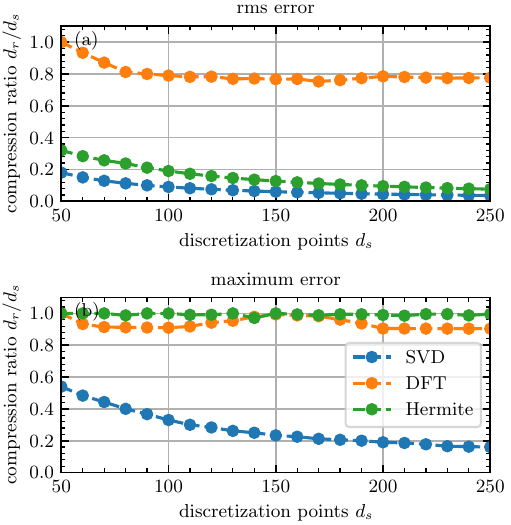}
\caption{Minimum compression ratio $d_r / d_s $ to achieve a reconstruction error score below $0.01$ as a function of the number of discretization points $d_s$. Blue, orange and green lines denote the SVD, DFT, and Hermite dimensionality reduction schemes. The error score adds up the mean and one standard deviation from a ten-fold cross-validation procedure of the training data. To evaluate the individual trajectories, both the rms error and the maximum absolute error are utilized and the results are presented in (a) and (b), respectively.
 }
\label{fig:dimensionality_compression}
\end{figure}

In order to benchmark the three dimensionality reduction schemes, we determine the minimum required compression ratio $d_r/d_s$ to achieve a reconstruction error score below $0.01$. For that purpose, we simulate training data sets (s. \cref{sec:training_data}) with an increasing number of discretization points $d_s \in \{50,60,\dots,250\}$, but identically constructed initial conditions.

To evaluate the quality of the reconstruction, we use a ten-fold cross-validation procedure (s. \cref{sec:benchmarking}) to obtain reconstruction errors for all individual trajectories. The error score is then computed as the mean plus one standard deviation in order to penalize large spreads in the individual reconstruction errors.
\Cref{fig:dimensionality_compression} shows the results for both the rms error (a) and the maximum absolute error (b) (s. \cref{sec:benchmarking}) as a function of the number of discretization points $d_s$. Blue, orange, and green denote the SVD, DFT, and Hermite reduction schemes. 

With respect to the rms error, the DFT reduction scheme performs rather poorly. For $d_s = 50$, all modes are required to achieve the desired error score. For an increasing number of discretization points $d_s$, the minimum compression ratio slightly improves but then plateaus around $d_r/d_s \approx 0.8$ for $d_s \gtrsim 100$. The SVD and Hermite schemes, on the other hand, offer much better error scores and exhibit a $\propto 1/d_s$ behavior. Specifically, the SVD scheme requires nine modes and the Hermite scheme requires 19 modes to achieve the error score target within the discretization point number range $80 \gtrsim d_s \geq 250$. Hence, the SVD reduction scheme performs better by a factor of two than the Hermite scheme with respect the rms error.

Those results, however, change when considering the maximum absolute error. In that case, both the DFT and the Hermite reduction scheme perform badly and exhibits minimum compression ratios close to one for all discretization point numbers $d_s$. The SVD reduction scheme, on the other hand, requires a compression ratio of $d_r/d_s = 0.52$ at $d_s = 50$ and improves to a compression rate of $d_r/d_s = 0.16$ at $d_s = 250$. 

We attribute the performance differences among the three different dimensionality reduction schemes to their specificity to the examined data. The DFT is naturally best suited for periodic and harmonic signals: two properties, which the considered electron distributions do not fulfill. The Hermite scheme achieves much better rms errors, since it is designed to describe electron distributions close to a quasi equilibrium. This condition is fulfilled for many system states of the considered trajectories. However, the Hermite reduction scheme fails to approximate the Gaussian initial states as well as the fringes at intermediate states well (c. \cref{fig:dynamics}). Hence, it only scores poorly with respect to maximum absolute error. The SVD mitigates this issue by being entirely data-driven and thus adapted to the initial, intermediate, and final system states along the nonlinear transients. Note, however, that the superior performance of the SVD reduction scheme is only achieved by providing a suitable training data-set. System states that are not well represented by the training data may suffer from strong reconstruction artifacts. A further discussion of the SVD reduction scheme applied to an example trajectory is presented in \cref{sec:SVD}.

The fact that the required compression ratio of the SVD based approach improves with an increasing number of discretization points demonstrates that the transient dynamics indeed exhibit low dimensional patterns. Hence, we choose the SVD approach to build the nonlinear autoregressive reduced order model.

We further would like to highlight that we only presented three dimensionality reduction schemes, which are all truncated linear transformations. Recent research on auto-encoder deep neural networks has produced promising and exciting results for low dimensionality latent representations of complex dynamical systems \cite{MIL02b,EIV20,EIV22}. Such approaches, however, come with expensive training and computational costs.

\subsection{Forecasting with polynomial features} \label{sec:poly}

Having chosen the appropriate dimensionality reduction scheme leaves us with designing the nonlinear transformation $\mathbf{f}_n = \mathbf{f} (\widetilde{\mathbf{r}}_n,\dots,\widetilde{\mathbf{r}}_{n-l+1})$ alongside a suitable feature scaling scheme.
In the following, we propose, evaluate, and discuss two transformations $\mathbf{f}(\cdot)$. The first one is based upon polynomial features and the second one generates its nonlinear features via an extreme learning machine (ELM).

In both cases, we first construct a linear feature vector by concatenating the past $\ell$ system states $\widetilde{\mathbf{r}}$
\begin{align}
    \mathbf{f}_n^\rm{lin} = \widetilde{\mathbf{r}}_n \oplus \widetilde{\mathbf{r}}_{n-1} \oplus \dots \oplus \widetilde{\mathbf{r}}_{n-\ell+1},
\end{align}
where $\oplus$ denotes the concatenation operator. This corresponds to the VAR$(\ell)$ state. For a reduced dimensionality $d_r$ of the state vectors $\widetilde{\mathbf{r}}_n$, the dimension of the linear feature vector $\mathbf{f}_n^\rm{lin}$ thus becomes $d_r \ell$.

In the case of the polynomial features transformation, we chose a degree $p$ and construct the feature vector according to
\begin{align}
    \mathbf{f}_n = 1 \oplus \mathbf{f}_n^\rm{lin} \oplus \rm{M}^{(2)}(\mathbf{f}_n^\rm{lin}) \oplus \dots \oplus \rm{M}^{(p)}(\mathbf{f}_n^\rm{lin}), \label{eq:poly_feature_vector}
\end{align}
where the operator $\rm{M}^{(p)}(\cdot)$ generates a vector, which contains all unique monomials of the order $p$ from the elements of a given input vector. The linear feature vector $\mathbf{f}_n^\rm{lin}$ can be understood as the first-order transformation and the constant element $1$ (bias/intercept term) as the zeroth-order transformation.
This transformation represents a truncated discrete Volterra series \cite{BOY85,FRA06a}, where the regression weights $w_{mn}$, which are to be estimated, correspond to the discrete-time Volterra kernels \cite{KOR88,MAT00a}. Note that the Volterra series is truncated both in time (only the past $\ell$ inputs are considered) and in the polynomial order $p$. On the one hand, this approach has the advantage of interpretable and intuitive features, but on the other hand, it requires the Volterra series to quickly converge in order to be computationally tractable. In particular, each order $p$ adds $\propto (d_r \ell)^p $ elements to the feature vector. In practice, however, low orders $p$ have often proved to be sufficient \cite{PYL21,GAU21b}. This, nonetheless, highlights that a small reduced state dimension $d_r$ is greatly desirable to minimize the dimension of the feature vector $\mathbf{f}_n$. 

To complete the model, we lastly specify the feature scaling scheme. This component determines the magnitude of the inputs, which the nonlinear transformation receives, and thereby critically controls its response. Since the polynomial features are unbounded, i.e., do not saturate, we want to avoid large outliers and scale each feature by the range, which we observe in the training data. Specifically, the $m$-th feature of the reduced state $\mathbf{r}_n$ is transformed according to
\begin{align}
    \widetilde{r}_{nm} = r_s \left[ \frac{ r_{nm} }{ \max_{j} \{r_{jm}\} - \min_j \{r_{jm}\} } - \frac{1}{2} \right].
\end{align}
The parameter $r_s$ determines the output range, to which the features are scaled, $j$ enumerates all system states in the training data, and the last term ensures that the scaled features are approximately symmetric with respect to the origin. We refer to this approach as feature normalization.

The entire model, which contains the SVD dimensionality reduction, the feature normalization, the polynomial features transformation, and the regularized least-squares optimization then contains the following hyper parameters: The number of past time steps $\ell$, the polynomial order $p$, the reduced state dimension $d_r$, the feature scaling range $r_s$, and the regularization parameter $\alpha$. In the following, we will explore the impact of the last three hyper parameters onto the model's prediction capabilities. The first two are kept constant at $\ell = 2$ and $p = 2$, since larger values offer almost no performance improvements for the considered data but increase the computational costs. Nevertheless, these two parameters should be optimized in the context of new data, where larger values of $\ell$ and $p$ correspond to systems with longer correlations and stronger nonlinearities, respectively.

\begin{figure}[htbp]
\centering
\includegraphics[width=\linewidth]{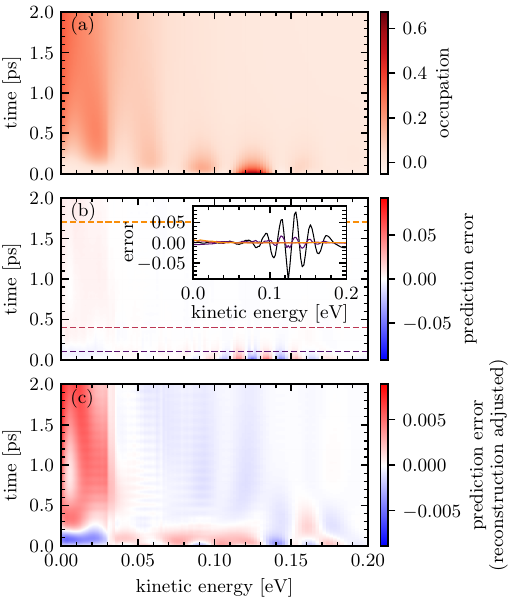}
\caption{Demonstration of the nonlinear autoregressive model with SVD-based dimensionality compression and polynomial features up to second order. (a) Example trajectory with the same initial state as in \cref{fig:dynamics}. (b) Prediction error plotted as the difference between the predicted trajectory $f_k^\rm{pred}$ and the ground truth $f_k$. The inset shows the error at $t 
\in \{0,0.1,0.4,1.7\}$\,ps (indicated by dashed horizontal lines). rms error $\epsilon_\rm{rms} = 0.0034$ and the maximum error $\epsilon_\rm{max} = 0.0926$. (c) Prediction error with respect to the reduced and reconstructed trajectory. rms error $\epsilon_\rm{rms} = 0.0022$ and the maximum error $\epsilon_\rm{max} = 0.0072$. 
Parameters: $d_s = 200$, $d_r = 20$, $\ell=2$, $r_s = 0.1$, $\alpha = 10^{-3.5}$.
 }
\label{fig:autoregression_example}
\end{figure}

To start off with, we want to demonstrate the forecasting performance of a well tuned model. For that purpose, as well as the following discussions, we simulate the electron-phonon dynamics with $d_s = 200$ discretization points. \Cref{fig:autoregression_example} then presents the results, which have been obtained with the identical initial conditions as in \cref{fig:dynamics}. The model has been trained on the complete training data set (s.\cref{sec:training_data}) and the hyper parameters are given in the figure caption.

Panel (a) plots the predicted transient electron dynamics, which visually appear very similar to the ground truth obtained from the simulation shown in \cref{fig:dynamics}. To highlight the small deviations, we plot the differences in the electron distribution $f^\rm{pred}(\epsilon_k,t) - f(\epsilon_k,t)$ in \cref{fig:autoregression_example}\,(b), where positive deviations are indicated by red colors and negative deviations by blue colors. The inset furthermore shows the error at the times $t \in \{0,0.1,0.4,1.7\}$\,ps, which are indicated by dashed horizontal lines in the main panel.

The largest absolute deviations are found at the bottom, i.e., in the beginning of the transient dynamics, where the distribution is still dominated by the Gaussian initial conditions. In particular, the forecasted system state periodically (in k-space) over and undershoots the true dynamics (black line in the inset), which is caused by a worst-case performance of the SVD dimensionality reduction scheme (s. \cref{sec:SVD}). The initial state consequently produces the maximum error $\epsilon_\rm{max} = 0.0926$.
However, once the electron distribution transitions to the intermediate and final states, its characteristic features broaden, which facilitates a much better reconstruction via the truncated SVD. Consequently, the absolute errors significantly reduce. For times $t \gtrsim 0.2\u{ps}$ the errors decrease below $ | f^\rm{pred}(\epsilon_k,t) - f(\epsilon_k,t)| = 0.01$.
A careful further inspection of \cref{fig:autoregression_example}\,(b) reveals that the most dominant prediction errors for times $t \gtrsim 0.2\u{ps}$ are found at kinetic energies $\epsilon_k \lesssim 0.15\u{eV}$. Here, the prediction overestimates the true dynamics as indicated by the weakly red colors on the left side. This also drives the root-mean-square (rms) error, which amounts to $\epsilon_\rm{rms} = 0.0034$.

To better understand the origin of the observed prediction errors, we moreover plot the prediction error with respect to the reconstructed ground truth $f^\rm{pred}(\epsilon_k,t) - f^\rm{recon}(\epsilon_k,t)$ in \cref{fig:autoregression_example}\,(c). This way, we can separate the reconstruction error and highlight the autoregression error. First and foremost, we find that the observed errors are reduced by approximately one order of magnitude when compared to the full error. The maximum error now only amounts to $\epsilon_\rm{max} = 0.0072$. The distribution of errors, however, is somewhat similar to the full error. At initial times $t \lesssim 0.3\u{ps}$, where nonlinear physics are the strongest, the regression both over and undershoots its targets. At later times, the errors are again most pronounced for small kinetic energies $\epsilon_k \lesssim 0.15\u{eV}$ as highlighted by the red stripe on the left. From this, we can conclude that the prediction error with respect to the true dynamics in this region is mostly caused by regression errors and not by reconstruction errors from the SVD scheme. It is further noteworthy, that the rms error yields $\epsilon_\rm{rms} = 0.0022$ with respect to the reconstructed dynamics. The comparison to $\epsilon_\rm{rms} = 0.0034$ from the true dynamics then reveals that the larger part is caused by regression errors. The maximum error, on the other hand, is mostly driven by the reconstruction error.
Given those insights, we will, nevertheless, be using the prediction error with respect to the true dynamics for further performance benchmarking, since we consider the dimensionality reduction stage as an integral part of the model.

\begin{figure}[htbp]
\centering
\includegraphics[width=\linewidth]{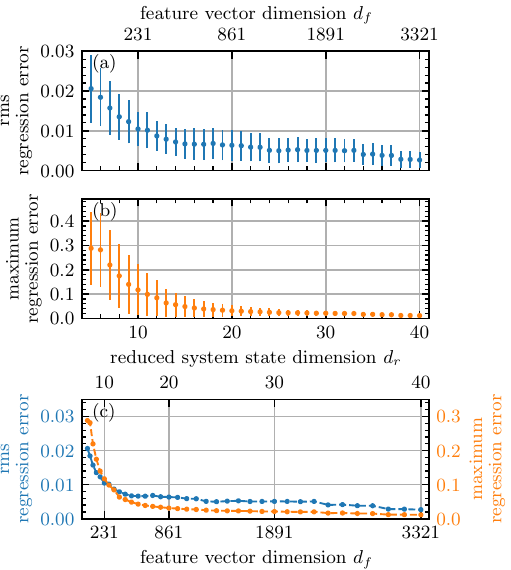}
\caption{Regression error as a function of the reduced dimensionality $d_r$. rms error (a) and maximum error (b). The errorbars indicate one standard deviation. Panel (c) plots the mean of both error norms as a function of the feature vector dimension $d_f$. The regularization parameter $\alpha$ is individually optimized for each $d_r$. Parameters: $\ell=2$, $p=2$, $d_s = 200$, and $r_s = 0.1$.
 }
\label{fig:poly_rdim}
\end{figure}

Having inspected the autoregression example presented in \cref{fig:autoregression_example}, we proceed to
study the impact of the reduced system state dimension $d_r$, which has previously been set to $d_r=20$. 
Unlike the other model hyper parameters, the reduced system state dimension not only affects the prediction accuracy, but also, critically, the computational costs. For that purpose, we set the feature normalization range to $r_s = 0.1$ and automatically optimize the regularization strength $\alpha$ for each scanned reduced dimension $d_r$. 
The results are obtained using a ten-fold cross-validation procedure (s. \cref{sec:benchmarking}) and are presented in \cref{fig:poly_rdim}, where (a) plots the rms error and (b) the maximum error. The error bars indicate one standard deviation. The dimension of the feature vector $\mathbf{f}_n$, which is given by
\begin{align}
    d_f = 1 + d_r\ell + d_r\ell(d_r\ell + 1)/2
\end{align}
for polynomial features up to second order, is further indicated on top of panel (a). 

We observe, that both error norms decrease with an increasing reduced state dimension, however, both with specific nuances. The rms error exhibits a relative plateau between around $15 \lesssim d_r \lesssim 32$, where both the mean and the standard deviation only marginally reduce. 
The maximum error, on the other hand, does not exhibit such a plateau. Both the mean and the standard deviation of the individual errors monotonically decrease with the reduced dimension $d_r$. We attribute this behavior to a better reconstruction ability of the full dynamics from a higher dimensional reduced state. As we have seen before, this especially benefits the distinct features of the early stage transients, which drive the maximum error. 
Since the smaller errors at the early stages of the transient also improves the rms error, we conclude that those play a minor role and the observed plateau arises due to dominating autoregression errors.

To gain practical insights from data, we relate the autoregression error to the computational costs of the model. For that purpose, \cref{fig:poly_rdim}\,(c) plots the mean rms error (blue) and the maximum error (orange) as a function of feature vector dimension. The reduced system dimension $d_r$ is indicated on the top. Note that we choose the feature vector size $d_f$, since it provides both a measure for computation time of the feature vector \cref{eq:poly_feature_vector} itself and the matrix multiplication of the autoregression step \cref{eq:autoregression_step}, and the memory requirements of the feature matrix \cref{eq:feature_matrix}, which is required for the model training procedure.

\begin{figure}[htbp]
\centering
\includegraphics[width=\linewidth]{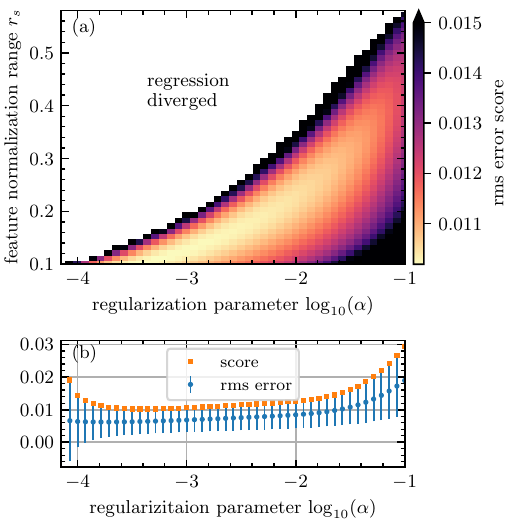}
\caption{Hyperparameter optimization of the nonlinear autoregressive model with polynomial features up to second order: The Tikhonov regularization parameter $\alpha$ and the feature scaling range $r_s$ are tuned to minimize the ten-fold cross-validated rms score. The rms error score is defined as the mean regression error plus one standard deviation. (a) Plots the color coded results and (b) a scan along $r_s = 0.1$, where blue denotes the regression error and orange the score. Parameters: $d_s = 200$, $d_r = 20$.
 }
\label{fig:poly_opt}
\end{figure}

In this representation, both curves clearly show elbows between $231 \lesssim d_f \lesssim 861$ ($10 \lesssim d_r \lesssim 20$) and exhibit diminishing returns for $d_f \gtrsim 861$ ($ d_r \gtrsim 20$). We therefore conclude that models with $d_r \in [10,20]$ represent the Pareto optimum \cite{BRU22}, i.e., they optimally balance small autoregression errors and model complexity. For this very reason, we have chosen the reduced dimension $d_r=20$ for the results presented in \cref{fig:autoregression_example} and \cref{fig:poly_opt}.

% Having inspected the autoregression example presented in \cref{fig:autoregression_example}, we proceed to 
% In this section, we
Lastly, we systematically study the impact of the regularization parameter $\alpha$ and the feature normalization range $r_s$. On that account, we define the autoregression error score as the mean plus one standard deviation of the rms error obtained from a ten-fold cross validation procedure (s. \cref{sec:benchmarking}) applied to the training data set (s. \cref{sec:training_data}). We specifically incorporate the standard deviation to encourage similar autoregression performances across different trajectories. The resulting map is color coded and presented in \cref{fig:poly_opt}\,(a).

The likely most striking fact is that the regression error score diverges in the upper left triangle (white color), i.e., for weak regularization parameters and/or large feature normalization ranges. This implies that the forecasting of at least one trajectory failed and diverged, which renders the associated hyper parameter set $(\alpha,r_s)$ unsuitable for general applications.
Stable operation, on the other hand, can be observed in the lower right triangle, i.e., at strong regularization parameters and/or small feature normalization ranges. 
The best scores are indicated by bright and the worst scores by dark colors. Note, however, that almost all non-divergent scores are found in the rather small range from $\approx 0.01$ to $\approx 0.015$. 

Quite importantly, the hyper parameter combinations, which offer the best error scores, can be safely chosen, since they exhibit a comfortable margin to the boundary of stability.
Among the best error scores is the configuration $\alpha = 10^{-3.5}$ and $r_s = 0.1$, which was used for the example presented in \cref{fig:autoregression_example} and produces the excellent score $0.0063 + 0.0039 = 0.0102$. Its autoregression error statistics are further discussed in \cref{sec:error_statistics}.

The structure of the regression error scores indicates that the regularization strength $\alpha$ and feature normalization range $r_s$ have a similar effect, even though they act differently on the model. 
The feature normalization range $r_s$ controls the magnitude of the inputs and thereby the response of the polynomial features transformation. On the one hand, large inputs destabilize the model due to their unbounded transformation (the unique monomials) and eventually cause the autoregression to diverge. On the other hand, small inputs quickly become negligible even for second-order monomials and thereby render the model approximately linear. 
The fact that feature normalization ranges $r_s \in [0.1,0.2]$ provide the best model performances thus indicates that the second-order monomials can be interpreted as a small but relevant correction to a linear model. It furthermore explains the lack of performance improvements observed for higher-order transformations ($p > 2$).

On the contrary, the regularization strength $\alpha$ only indirectly affects the nonlinear response via errors that are accumulated during the repeated self-injection of the predicted state as a new input (s. \cref{fig:model}). Tikhonov regularization is a well established \cite{VOG02,PRE07,BRU22} approach to improve the out-of-sample prediction for regression problems with multicollinear features. This turns out to be especially relevant in our case, since the polynomial features of our transformation are not orthogonal and thus create approximately multicollinear features by construction. Hence, for each feature normalization range $r_s$, a specific regularization strength $\alpha$ best balances the stability (strong $\alpha$) and the accuracy (weak $\alpha$) of the autoregressive model.

The impact of the regularization strength is further illustrated in \cref{fig:poly_opt}\,(b), which presents the regression error as function of the regularization parameter $\alpha$ for the fixed feature normalization range $r_s = 0.1$. Blue circles and error bars indicate the mean rms error and its standard deviation and orange squares the autoregression error score as it is used in \cref{fig:poly_opt}\,(a). 
Increasing the regularization strength starting from $\alpha = 10^{-4.1}$, we observe a decreasing error score to a global minimum around $\alpha \approx 10^{-3.5}$. Notably, the improvement is mostly due to a decreasing standard deviation, which we attribute to an increase in stability as discussed above.
Further increasing the regularization strength then leads to a monotonically growing mean regression error, which we attribute to a lack of specificity and adaption of the regression weights due to the stronger regularization constraints. Similarly, the standard deviation also grows such that the corresponding relative error remains approximately constant. A qualitatively identical behavior can also be observed for other feature normalization ranges $r_s$, as can be seen in \cref{fig:poly_opt}\,(a).

In summary, the nonlinear autoregressive reduced order model built upon polynomial features offers an excellent forecasting performance and interpretability due to its connection to discrete Volterra series \cite{BOY85,FRA06a}. Moreover, the individual features can be easily computed via the upper(lower) triangle of the linear feature vector's $p$th-order tensor product. On the downside, however, one must be very careful with the feature scaling and regression weight regularization, in order to avoid diverging forecasts. Furthermore, the feature vector size scales with the order of the polynomial features transformation, which can become restrictive even at the second order ($p=2$) for large linear feature vectors.

\subsection{Forecasting with ELM features} \label{sec:ELM}

With the intend to overcome some of the shortcomings of the polynomial features transformation, we further propose and investigate a nonlinear transformation based on an extreme learning machine (ELM) \cite{HUA04c}. ELMs refer to feed-forward artificial neural networks, where only the linear output layer, but not the internal hidden layers, are trained. This allows for a simple one-step learning, which can be orders of magnitude faster than the gradient-based methods employed for the complete training of artificial neural networks, while also keeping the required training data sets small \cite{HUA06,PYL21}. Despite this restriction, ELMs have been shown to have universal approximation capabilities for continuous regression tasks \cite{HUA06a,HUA11a}. Since they are not part of the training procedure, the internal parameters are typically drawn randomly from suitable distributions. The internal layers therefore perform a random projection of the input data into a nonlinear feature space, from which the output is generated via the trained linear mapping.

In the case of the ELM transformation, we construct the feature vector according to
\begin{align}
    \mathbf{f}_n = 1 \oplus \mathbf{f}_n^\rm{lin} \oplus \mathbf{f}^\rm{ELM}( \mathbf{f}_n^\rm{lin} ).
\end{align}
The first two vectors represent the bias term and the linear features and the last vector represents the nonlinear features generated by an ELM. Specifically, the function $\mathbf{f}^\rm{ELM}(\cdot)$ is given by
\begin{align}
    \mathbf{f}^\rm{ELM}( \mathbf{f}_n^\rm{lin} ) = \phi \left( \mathbf{W}_\rm{ELM} \mathbf{f}_n^\rm{lin} + \boldsymbol{\beta} \right),
\end{align}
where a nonlinear activation function $\phi$ acts on each element of the input vector, which is generated by the weight matrix $\mathbf{W}_\rm{ELM} \in \mathbb{R}^{L \times d_f}$, the linear feature vector $\mathbf{f}_n^\rm{lin}$, and the bias vector $\boldsymbol{\beta} \in \mathbb{R}^{L \times d_f}$. 
Hence, each nonlinear feature is given by a nonlinearly transformed linear combination of all linear features. 
This nonlinear transformation can be represented by a fully connected feed-forward network with one hidden layer, to which a nonlinear activation function is applied. 
The matrix $\mathbf{W}_\rm{ELM}$ further defines the number of neurons (nodes) $L$ in the hidden layer via its shape $\mathbb{R}^{L \times d_f}$.
This fact constitutes a major advantage of the ELM approach compared to the polynomial features: The size of the feature vector, which is relevant for the computational costs, does not depend on the order of the nonlinearity and can be explicitly controlled.

For our purposes, we use the hyperbolic tangent $\tanh(\cdot)$ as the nonlinear activation function $\phi(\cdot)$. Note that this nonlinearity behaves fundamentally different than the previously used polynomial transformation (s. \cref{sec:poly}). For small inputs $|u| \ll 1$, the hyperbolic tangent behaves linearly; for intermediate inputs $|u| \sim 1$, the hyperbolic tangent is strongly nonlinear; and for large inputs $|u| \gg 1$, the hyperbolic tangent saturates.
Hence, we can tune the strength of the nonlinearity by controlling the input $u$, i.e., the magnitude of the individual features, without ever risking a diverging response.

On that account, we standardize the individual features by subtracting their mean $\mu_m$ and scaling them by their standard deviation $\sigma_m$:
\begin{align}
    \widetilde{r}_{nm} =  \frac{ r_{nm} - \mu_m }{ \sigma_m }. \label{eq:ELM_scaling}
\end{align}
Coordinated with that, we draw the weights $w_{mn}^\rm{ELM}$ from a normal distribution $\mathcal{N}^{\mathbf{W}}(0,d_f^{-1})$ with zero mean and the variance given be the inverse feature vector dimension $d_f$. Given the standardization of the features, the scaling of the variance ensures that the expected magnitude of the inputs $u_{nl} = \sum_m w_{lm}^\rm{ELM} f^\rm{lin}_{nm}$ is of the order $\approx 1$.
This construction has been shown to optimally harness the $\tanh$ nonlinaerity and produce the best regression results \cite{AKU15}.
Similarly, the biases $\beta_m$ are drawn from the uniform distribution $\mathcal{U}^\mathbf{\beta}(-1.0,1.0)$.

The insight that the presented combination of feature scaling and random weights already optimizes the regression potential of the ELM features therefore relieves the human operator from the obligation to optimize any of the related parameters. This simplification thus facilitates an easier implementation of the nonlinear autoregressive reduced order model.

\begin{figure}[htbp]
\centering
\includegraphics[width=\linewidth]{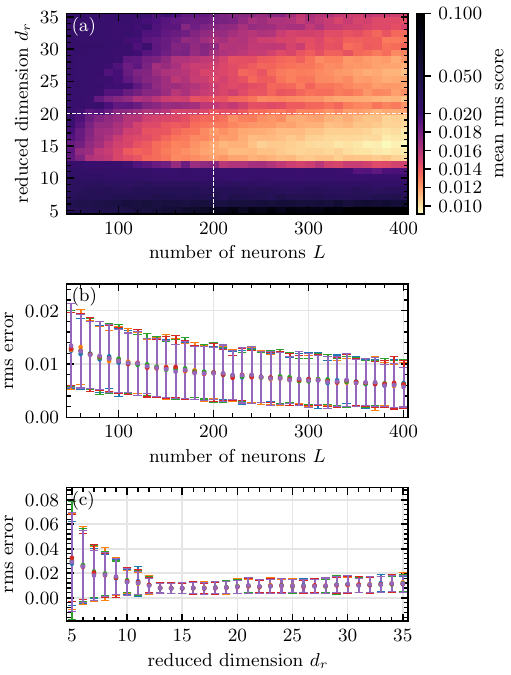}
\caption{Hyperparameter optimization of the nonlinear autoregressive model with ELM features in the parameter space spanned by the number of ELM neurons $L$ and the reduced dimensionality $d_r$.
Panel (a) plots the mean rms score, which is evaluated as the regression error plus one standard deviation averaged over five different sets of random ELM parameters $(\mathbf{W}^\rm{ELM},\mathbf{\beta})$. The regularization parameter $\alpha$ is optimized for each parameter combination. Panel (b) and (c) plot one-dimensional slices along $d_r = 20$ and $L = 200$ (indicated by white dashed lines in (a)), where different colors denote the five ELM parameter sets.
Other parameters: $\ell=2$ and $d_s = 200$.
 }
\label{fig:ELM_opt}
\end{figure}

We forgo the presentation of the ELM based autoregression applied to the example trajectory (s. \cref{fig:dynamics}), since the results turn out to be very similar to polynomial features based autoregression (s. \cref{fig:autoregression_example}).
In order to investigate the autoregression performance of ELM transformation, we evaluate the rms error in the model hyper-parameter plane spanned by the reduced system dimension $d_r$ and the number of ELM neurons $L$. We optimize the regularization parameter $\alpha$ independently via a simple linear search for each parameter combination. To account for the randomly drawn ELM parameters $(\mathbf{W}^\rm{ELM},\mathbf{\beta})$, we consider five different realizations. \Cref{fig:ELM_opt}\,(a) plots the color-coded mean rms score, which adds the mean regression error and one standard deviation and averages the result over the five different ELM parameter sets.

For an increasing number of ELM neurons $L$, we observe improving autoregression scores, irregardless of the reduced system dimension $d_r$. This is in accordance with the ELM literature \cite{HUA06a,HUA11a}, where more ELM neurons increase the nonlinear feature space and thus improve the ELM's approximation capabilities. 
Scanning along the reduced system dimension $d_r$, however, yields a different behavior: For an increasing reduced system dimension, the autoregression score initially decreases rapidly , then reaches a global minimum around $d_r \sim 15$, and finally slowly increases again. It is further important to note that the minimum itself shifts to a larger reduced dimensions $d_r$ with an increasing number of neurons $L$. Specifically, from $d_r^\rm{min} = 13$ at $L = 50$ to $d_r^\rm{min} = 15$ at $L=400$.
We interpret this result as follows: If the reduced state dimension is increased, the nonlinear feature space must be increased in order to accommodate the additional information and to reliably generate suitable features, upon which the regression step is build. Moreover, the additional information, which is provided by the additional features dimensions, has a diminishing relevance for the reconstruction of the full system state (proportional to its corresponding singular value $\sigma_l$, s. \cref{fig:SVD}, note the kink at $l=14$ in (a1)). Hence, keeping the number of neurons constant and increasing the reduced state dimension might wash out the more relevant contributions in the first few components via the random linear combination with the ELM weight matrix $\mathbf{W}_\rm{ELM}$.

We further illustrate those findings by showing one dimensional scans along the number of ELM neurons $L$ for $d_r = 20$ in \cref{fig:ELM_opt}\,(b) and along the reduced dimension $d_r$ for $L = 200$ in \cref{fig:ELM_opt}\,(c). These scans are indicated by white dashed lines in \cref{fig:ELM_opt}\,(a). However, unlike \cref{fig:ELM_opt}\,(a), we plot the rms error with one standard deviation indicated by errorbars and represent the five different ELM realizations by different colors. 
Both (b) and (c) reveal that the changes in the rms score, as they are portrayed in (a), are produced by roughly equal changes of the mean and the standard deviation. Thus, the error score minimum at $d_r = 14$ both represents the smallest mean and the smallest standard deviation.
Moreover, the separate representation allows to eyeball the impact of the randomly drawn ELM weights $\mathbf{W}^\rm{ELM}$. For all investigated hyper parameters, the variation of both the mean and the standard deviation is well within the the corresponding error ranges. Hence, we conclude that an ELM based model can be initially implemented and optimized without giving extra attention to the ELM realization. Nonetheless, the autoregression performance can be maximized by generating multiple ELMs and picking the best.

Furthermore, we report that the optimization of the regularization parameter $\alpha$ is of relatively small importance for the ELM based model. Unlike the polynomial features based model (s. \cref{sec:poly}), we only obtain diverging trajectories for very small reduced system dimensions $d < 8$ and almost negligible regularization strengths $\alpha < 10^{-5}$. Moreover, for reduced system dimensions $d_r \geq 12$, the optimizations of regularization strength typically yields improvements of only a few percents. For practical and fast implementations of the model, the regularization parameter can therefore be safely set to intermediate values such as $\alpha = 10^{-2}$ and ignored afterwards.

In summary, the ELM based implementation of the nonlinear autoregressive reduced order model yields very competitive results with only few hyper parameters to be optimized. The unique hyper parameter, i.e. the number of neurons $L$, can be furthermore understood as an advantage, since it directly affects the computational performance and can be explicitly set. Moreover, the ELM approach shows a high resilience against diverging trajectories, which we attribute to the saturating tanh nonlinearity.
On the downside, the evaluation of the tanh nonlinearity is computationally more expensive than the monomials. Furthermore, the slight stochasticity of the forecasting performance due to the randomly drawn weights inhibits the usage of hyper-parameter optimization algorithms, e.g. Bayesian optimization, which require smooth goal functions.

\subsection{Training data set size dependence}

\begin{figure}[htbp]
\centering
\includegraphics[width=\linewidth]{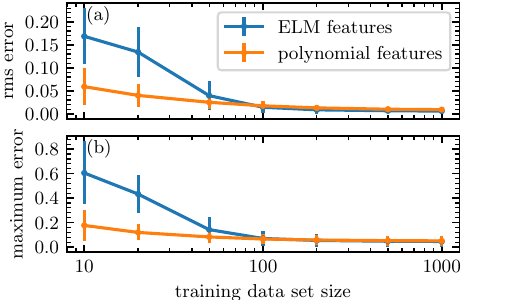}
\caption{Autoregression error as function of the size of the training data set. For sizes $<1000$ the complete data set with $n=1000$ is subsampled to generate error scores for all trajectories via a ten-fold cross-validation procedure for each subset. Autoregression is performed with SVD based dimensionality reduction and ELM features (blue) and polynomial features (orange). Parameters: $d_s = 200$, $\ell=2$, $d_r = 15$, and $\alpha = 10^{-2}$. ELM parameters: $L=400$. Polynomial features parameters: $p=2$ and $r_s = 0.1$.
 }
\label{fig:training_data_dependence}
\end{figure}

Lastly, we study the relation between the autoregression error and the training data set size. For that purpose, we subsample the complete training data set with $n=1000$ (s. \cref{sec:training_data}) into chunks with the sizes $n \in \{10,20,50,100,200,500\}$. Both for a model with ELM and a model with polynomial features, we then obtain a regression error for each trajectory from the complete set by performing a ten-fold cross-validation procedure on each of the subsets (s. \cref{sec:benchmarking}). 
Inferring the optimal ELM parameters from \cref{fig:ELM_opt}, we set the relevant model parameters to $d_r = 15$ and $L=400$, which yields a feature vector with the size $d_f = 431$. Similarly, the polynomial features model is operated with $d_r = 15$, which yields a feature vector with the dimension $d_f = 496$. Hence, both models have roughly the same number of regression weights $w_{nm}$ that must be optimized by exposing them to the training data. The regularization parameter is set to $\alpha = 10^{-2}$ for both models, which was required to ensure that no forecasts with the polynomial features based model diverge for small training data set sizes.
The results in terms of the mean regression error and its corresponding standard deviation are presented in \cref{fig:training_data_dependence}, where (a) shows the rms error and (b) the maximum error. Blue denotes the model with ELM features and orange the model with polynomial features.

For both error scores and both models, we find that the regression error and its standard deviation monotonically decreases with the training data set size. However, both error scores clearly exhibiting diminishing returns, both in terms of the mean and the standard deviation, for training data set sizes beyond $n \sim 100$. This behavior is especially prominent for the model with ELM features, whose autoregression performance degrades much stronger for training data set sizes below $n\sim 100$. Note, however, that the ELM features based model also achieves a slightly better performance for larger training data set sizes ($n \gtrsim 100$) than the polynomial features based model.
We attribute the advantage of the polynomial features based model at small training data set sizes to a better specificity of the transformation. I.e., the truncated Volterra series is a more 'natural' representation of nonlinear dynamics at hand, which facilitates a data efficient optimization of the regression weights $w_{nm}$ and a better generalization capabilities on unseen trajectories.
Moreover, we want to highlight that the diminishing returns effect is slightly stronger for the maximum regression error, which we attribute to the limits of the reconstruction error from the reduced order latent space.

In conclusion, we find that $n \sim 100$ training trajectories are sufficient to obtain a well trained model for the problem at hand, irregardless of the chosen nonlinear feature mapping. However, if only few training trajectories are available, the polynomial features transformation offers a better performance than the ELM features transformation.

\section{Discussion} \label{sec:discussion}

In this work, we have demonstrated a nonlinear autoregressive reduced-order model (NARROM) for the forecasting of nonlinear transient dynamics in a coupled electron-phonon system. This data-driven model is designed to be minimal and thus efficient, both in terms of the computational training and forecasting costs, in order to replace expensive simulations in multi-physics problems.
Our approach capitalizes on a dimensionality reduction scheme, which extracts an optimal set of low dimensional patterns from the transient dynamics. Those patterns constitute a reduced-order latent space, which facilitates a computationally efficient, yet accurate, approximation of the system state.
In particular, the reduced dimensionality helps to mitigate the curse of dimensionality. This becomes especially relevant, since the (reduced) system states are projected into a generally higher dimensional nonlinear feature space, before performing the regression step.

Our results show, that a dimensionality reduction scheme based on the singular value decomposition extracts an optimal truncated basis for the considered transient dynamics, and thereby easily outperforms established dimensionality reduction schemes based on Fourier modes and Gauss-Hermite functions.
Hence, we recommend a SVD based dimensionality reduction scheme almost unconditionally - the only caveat is the lack of a straight-forward physical interpretation of the extracted modes.
Regarding the nonlinear feature mapping, we must formulate a recommendation with some important nuances. We have demonstrated that both the polynomial features and the ELM features can yield excellent forecasting performances. 
On the one hand, the polynomial features are easily constructed and evaluated and provide an intuitive interpretation in terms of the well-established discrete Volterra series. On the downside, however, the feature space dimension scales with the largest polynomial degree and the features must be appropriately scaled and regression weights must be well regularized in order to avoid diverging forecasts.
On the other hand, the ELM features yield a model that is much less prone to diverging forecast and thus requires fewer hyper parameters to be optimized. This allows for a more straight forward and safe deployment of the model. On the downside, the utilized $\tanh$ nonlinearity is more expensive to evaluate than the polynomials and thus comparatively deteriorates the computational performance of an ELM based model.
This apparent tie between the two nonlinear transformations is broken by their response to different training data set sizes: Polynomial features offer a better performance for small data sets, but loose that advantage at larger data sets.
Hence, we recommend the use of the polynomial features transformation for small training data set sizes. The additional optimization of the critical model hyper parameters is facilitated by the inexpensive training due to the small training data set size. The optimization may further be automatized, e.g., by a Bayesian optimization algorithm. For larger training data set sizes, we recommend the ELM features transformation due to its easier deployment and superior stability.

\begin{figure}[htbp]
\centering
\includegraphics[width=\linewidth]{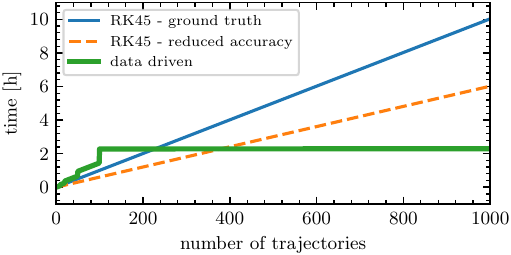}
\caption{Simulation time as a function of the number of trajectories. 
The blue line denotes the ground truth as obtained from the integration of the differential equations with an adaptive step-size Runge-Kutta algorithm. The dashed orange line denotes the simulation with reduced accuracy.
The thick green line denotes the data-driven approach, where the initial steps result from generating the training data and optimizing the model hyperparameters. The latter two are configured to yield an average maximum error of $\lesssim 0.06$.
 }
\label{fig:simulation_costs}
\end{figure}
Up to this point, we have abstained from the discussion of the computational demands of the nonlinear autoregressive reduced-order model in comparison to the direct integration of the equations of motion. The bare results of such a comparison are to be taken with a grain of salt, because both the runtime and the allocated memory critically depend on the actual implementation the equations of motion, the ODE solver, and the data-driven model. 
% It is therefore more sensible to discuss the computational complexity.
Nonetheless, we want to illustrate the performance figures, which we have achieved to highlight the qualitative trends.
With our implementation, the simulation of the training data set with $n=1000$ and $d_s =200$ takes about 10 hours on an Intel Core i7-8700 with six physical cores. 
In our test scenario, we decide that this time is unfeasible, but we are willing to accept and average maximum error of $\epsilon_\rm{max} \approx 0.06$ for the trajectories. 
The traditional approach would be to decrease the accuracy of the integration method.
With the data-driven approach, on the other hand, we first have to generate sufficient training data and then optimize the model hyperparameters. To be efficient, we simulate trajectories with quasi-logarithmic steps, i.e, $10,20,50,100,\dots$, and perform a hyperparameter optimization each time until the cross-validated error fulfills the threshold.

We presented the results in \cref{fig:simulation_costs}, where the simulation time is plotted as a function of number of trajectories. 
The blue line denotes the direct integration of the equations of motion \cref{eq:elec},\cref{eq:phon} with a forth-/fifth-order adaptive step-size Runge-Kutta algorithm and represents the ground truth.
The dashed orange line denotes the integration with the same method, but with reduced accuracy.
The thick green line denotes the data-driven approach, where we use ELM features, and fix the delay embedding to $\ell=2$ and the regularization to $\alpha = 10^{-2}$. The reduced dimension $r_d$ and number of Neurons $L$ are subject to an optimization, which we perform via a simple grid search with 121 sample points. 
The integration of the differential equations yields a linearly increasing simulation time, where the reduced accuracy integration produces a $\approx 66\%$ speed improvement. 
The data-driven approach, on the other hand, exhibits are more nuanced behavior, where the initial steps result from generating the training data and optimizing the model hyperparameters. A satisfactory model is obtained with $n=100$ training trajectories, which is consistent with \cref{fig:training_data_dependence}. With the trained model, the simulation time then grows linearly. However, it only amounts to 27 seconds for the remaining 900 trajectories and therefore becomes negligible and appears flat in the figure. The turning point, where the data-driven approach gains an advantage over the reduced-accuracy integration is then located at $n=380$ trajectories. 

Hence, the data-driven approach becomes a viable option for any application, where $n > 380$ trajectories are required. However, it truly becomes powerful and may save orders of magnitude in simulation time, if $n \gg  380$ trajectories are needed, because the costs of obtaining the trained model remain the same and the simulation of further trajectories is very inexpensive.
We attribute the enormous speed-up of forecasting individual trajectories with the trained model to three factors: 
Firstly, the data-driven model propagates the nonlinear system in the reduced-dimensionality latent space.
Secondly, the data-driven model can take nonlinear steps and therefore tolerates much coarser temporal discretizations. 
Lastly, the phonon system must not be explicitly propagated, but is implicitly included in the delay-embedding of the electron system. 

Moreover, the computational demands of directly simulating the coupled system scale with the square of the number of discretization points $d_s$. The computational costs of the data-driven model, on the other hand, only scale linearly with the number of discretization points (the dimensionality reduction and reconstruction stages). Thus, the data-driven model's advantage becomes even more pronounced for fine $k$-space discretizations.
Lastly, we want to highlight that most of the proposed data-driven methods build upon simple and well established linear algebra routines, which makes them easy to adapt and to implement, and computationally inexpensive to run.

Our results led us to conclude that the approximation of computationally expensive nonlinear dynamics by a minimal data-driven model (as opposed to, e.g., a deep artificial neural network) has the potential to greatly accelerate the simulation of multi-physics problems.
This may have profound implications for the further development of, among others, solid-state based opto-electronic devices.

\section{Acknowledgements}

We acknowledge fruitful discussions with Robert Salzwedel and Jonas Grumm (TU Berlin).\\
This work was funded by the Deutsche Forschungsgemeinschaft (DFG) through Project SE 3098/1, Project No. 432266622 (M.S.), SFB 951, Project No. 182087777 (D.K. and A.K.), and SFB 910 project B9 (F.K.).

\section{Author contributions}
S.M., F.K., and M.S. initiated and conceptualized the work. M.S., D.K. and A.K. worked on the microscopic derivation and the numerical implementation of electron-phonon scattering and parameter search. S.M. and F.K. implemented the nonlinear autoregressive reduced-order model. S.M. performed the simulations and numerical experiments. S.M. and M.S. drafted the manuscript. All authors discussed and edited the manuscript.

\section{Data Availability}
The data generated in this work can be generated by running the publicly available code as described in the code availability statement.

\section{Code Availability}
The simulation code and the regression code is available on GitHub under MIT license (\url{https://github.com/stmeinecke/derrom}) 
% \sm{and deposit once with doi to be cited}

\appendix

\section{Equations of Motion} \label{sec:eom}

In this section, we briefly discuss the coupled electron-phonon dynamics. To derive the corresponding equations of motion, we start with the parametrization of a microscopic Hamiltonian containing the dispersion of electrons and phonons as well as the interaction between them. The electron dispersion is treated in the parabolic approximation, with parameters taken from ab-initio calculations\cite{Kormanyos2015}. The dispersion of the acoustic phonons is treated in the Debye approximation with the velocity of sound taken from ab-initio calculations\cite{Li2013}. The dispersion of the optical phonons is treated in the Einstein approximation with parameters taken from ab-initio calculations\cite{Li2013}. To study the coupled electron phonon dynamics, our observables are the electron occupation $f_\mathbf{k} = \langle c^\dagger_\mathbf{k} c_\mathbf{k} \rangle$, with electron annihilation (creation) operators $c^{(\dagger)}_\mathbf{k}$ with the momentum $\mathbf{k}$ and the phonon occupation $n_\mathbf{q} = \langle b^{\dagger \alpha}_\mathbf{q}b^{\alpha}_\mathbf{q} \rangle$, with phonon annihilation (creation) operators $b^{(\dagger) \alpha}_\mathbf{q}$ with the phonon branch $\alpha$ and the phonon momentum $\mathbf{q}$. To derive an equation of motion for the electron occupation $f_\mathbf{k}$ and the phonon occupation $n_\mathbf{q}^\alpha$, we exploit the Heisenberg equation of motion. The upcoming hierarchy problem is treated in a correlation expansion and a second-order Born-Markov approximation\cite{butscher2007hot}. The resulting coupled electron-phonon Boltzmann scattering equations read
\begin{widetext}
\begin{align}
\partial_t f_\mathbf{k} &= \frac{2 \pi }{\hbar} \sum_{\mathbf{k'},\alpha,\pm} |g_{\mathbf{k} - \mathbf{k'}}|^2 \left(\frac{1}{2} \pm \frac{1}{2} + n_\mathbf{\mathbf{k} - \mathbf{k'}}^\alpha \right) \left( 1 - f_\mathbf{k} \right) f_\mathbf{k'} \delta (\epsilon_\mathbf{k} - \epsilon_\mathbf{k'} \pm \hbar \omega^\alpha_\mathbf{k - k'})\nonumber \\
&- \frac{2 \pi }{\hbar} \sum_{\mathbf{k'},\alpha,\pm} |g_{\mathbf{k} - \mathbf{k'}}|^2 \left(\frac{1}{2} \pm \frac{1}{2} + n_\mathbf{\mathbf{k} - \mathbf{k'}}^\alpha \right) \left( 1 - f_\mathbf{k'} \right) f_\mathbf{k} \delta (\epsilon_\mathbf{k} - \epsilon_\mathbf{k'} \mp \hbar \omega^\alpha_\mathbf{k - k'})\label{eq:elec} \\
\partial_t n_\mathbf{q}^\alpha &= \frac{2 \pi }{\hbar} |g_\mathbf{q}^\alpha|^2 \sum_\mathbf{k}  \left( 1 - f_\mathbf{k} \right) f_\mathbf{k+q} \left( 1 + n_\mathbf{q} \right) \delta (\epsilon_\mathbf{k} - \epsilon_\mathbf{k+q} + \hbar \omega^\alpha_\mathbf{q})\nonumber \\
&- \frac{2 \pi }{\hbar} |g_\mathbf{q}^\alpha|^2 \sum_\mathbf{k}  \left( 1 - f_\mathbf{k+q} \right) f_\mathbf{k} n_\mathbf{q}  \delta (\epsilon_\mathbf{k} - \epsilon_\mathbf{k+q} - \hbar \omega^\alpha_\mathbf{q}).\label{eq:phon}
\end{align}
\end{widetext}
Equation \ref{eq:elec} describes the temporal evolution of the electron occupation where the first term on the right-hand side accounts for in-scattering processes and the second line accounts for out-scattering processes. In both lines, the $+$ terms in the summation account for phonon emission processes and the $-$ terms account for phonon absorption processes. The rates depend on the electron-phonon coupling matrix element $g_\mathbf{k}$ with the momentum $\mathbf{k}$. The appearing delta functions ensure energy conservation during an electron-phonon scattering event. We treat the dispersion of electrons in the parabolic approximation $\epsilon_\mathbf{k} = \frac{\hbar^2 \mathbf{k}^2}{2 m_e}$, the dispersion of acoustic phonons in the Debye approximation $\hbar\omega_\mathbf{k}^a = c_{LA} |\mathbf{k}|$, and for optical phonons in the Einstein approximation $\hbar\omega_\mathbf{k}^a = \hbar \omega^o$.
Equation \ref{eq:phon} describes the temporal evolution of the phonon occupation. The first line on the right-hand side accounts for phonon emission processes and the second line accounts for phonon absorption processes. We numerically integrate the coupled system of equations of motions for the exemplary two-dimensional material MoSe$_2$.

For our evaluation we treat the appearing electron coupling elements $g_\mathbf{k}$ in the effective deformation potential approximation \onlinecite{Li2013}
\begin{equation}
g_q^{i} = \sqrt{\frac{\hbar}{2 \rho \Omega^i A}} V_q,
\end{equation}
with the effective mass density of the unit cell $\rho$ and the semiconductor area $A$. The effective deformation potential reads for acoustic phonon coupling $V_q=D_1 q$, and for optical phonons $V_q=D_0$. We take the parameters from reference \onlinecite{Jin2014} and list them in table \ref{tab_e_phon}.

\begin{table}[h!]
\centering
 \caption{Material paramters used in the calculation }
 \begin{tabular}{c|c|c}
   \hline
 $m_e$/$m_0$ & 0.5 & \onlinecite{Kormanyos2015} \\  
 $\bar \omega^o$/meV & 36 & \onlinecite{Jin2014}\\
 $c_{LA}$/(nm/fs) & 4.1 & \onlinecite{Jin2014}\\
 $D^{a}_1$/eV & 3.4 & \onlinecite{Jin2014}\\
 $D^{o}_0$/eV nm$^{-1}$ & 52 & \onlinecite{Jin2014}\\
 \end{tabular}\label{tab_e_phon}
\end{table}

\section{Benchmarking} \label{sec:benchmarking}

With the intention of forecasting transient dynamics over extended time periods in the closed-loop autonomous mode of the model (s. \cref{fig:model}), we want to evaluate the model performance for complete trajectories and not only for individual time steps. Given a test trajectory $\mathbf{S}$ and its forecasted approximation $\hat{\mathbf{S}}$, we use two different error scores for that purpose. Firstly, the root-mean-squared (rms) error
\begin{align}
    \epsilon_\rm{rms} = \sqrt{ \frac{1}{N} \sum_{m,n} \left( \hat{s}_{mn} - s_{mn} \right)^2 } = \sqrt{\frac{1}{N_\mathbf{S}}} \norm{\hat{\mathbf{S}} - \mathbf{S}}_\rm{F},
\end{align}
which corresponds to the Frobenius norm $\norm{\cdot}_\rm{F}$, which is normalized by the square root of the number of matrix elements $N_\mathbf{S}$. The normalization ensures that different discretizations of the dynamics produce comparable errors.
Secondly, the maximum error
\begin{align}
    \epsilon_\rm{max} = \max_{mn} \big| \hat{s}_{mn} - s_{mn} \big|,
\end{align}
which yields the element-wise maximum absolute difference between $\hat{\mathbf{S}}$ and $\mathbf{S}$. While the maximum error highlights the worst part of an approximation $\hat{\mathbf{S}}$, the rms error considers an average over all squared errors. However, large individual errors are still represented strongly due to the square. Moreover, the rms error directly relates via the Frobenius norm to the loss function \eqref{eq:lossfunction}, which is minimized to obtain the regression weights.

To benchmark a given model, we use a k-fold cross validation scheme, where we split the training data set (in terms of trajectories) into k equally sized folds. For each fold, we train the model on all the other folds and then score the trajectories contained in the selected fold. This way, we do not mix training and testing data but, nonetheless, obtain a score for each trajectory in the data set. Using the individual error scores, we can then compute the desired statistics, e.g., the mean score and the standard deviation of the scores.

\section{Training Data} \label{sec:training_data}

To generate a training data set, we simulate 1000 trajectories with varying initial electron distributions. As we want to consider the transient dynamics of the solid-state system shortly after the excitation by a short laser pulse, we use thermal distributions for the phonon system and a Gaussian distribution for the electron system. 
The mean, the standard deviation, and the maximum value of each Gaussian distribution are drawn from the continuous uniform distributions $\mathcal{U}^\rm{mean}(0.0\u{eV},0.175\u{eV})$, $\mathcal{U}^\rm{std}(0.005\u{eV},0.025\u{eV})$, and $\mathcal{U}^\rm{max}(0.5,0.99)$, respectively. The variations of the Gaussian parameters is attributed to laser pulses with different mean frequencies, pulse widths, and powers. The bounds of the uniform distributions are chosen to ensure strong nonlinearities in the resulting dynamics. 
Each initial condition is integrated for $2\u{ps}$ and then sampled every $50\u{fs}$ to yield sequences with 400 equidistant snapshots of the system state. The sampling interval has been chosen to yield visually smooth trajectories, without unnecessarily inflating the training data size. Tuning the sampling interval within a reasonable range has not been found to significantly impact the forecasting performance. 

\begin{figure}[htbp]
\centering
\includegraphics[width=\linewidth]{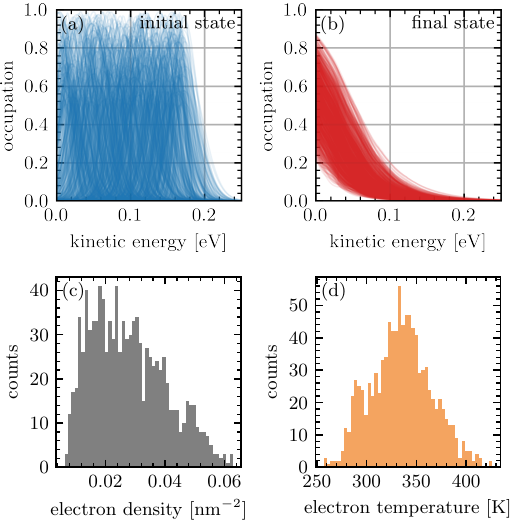}
\caption{Illustration of the training dataset with $n=1000$ trajectories. (a) Randomly drawn Gaussian initial states of the electron distribution. (b) Final states after $2$\,ps integration time. (c) Histogram of the electron densities. (d) Histogram of the electron temperature after $2$\,ps integration time. Parameters: $d_s = 200$.
 }
\label{fig:dataset}
\end{figure}

The resulting training data is illustrated in \cref{fig:dataset}, where (a) and (b) plot the initial electron distribution at $t = 0\u{ps}$ and the final electron distribution at $t=2\u{ps}$ for each of the 1000 trajectories. The randomly generated Gaussian initial conditions first cover a diverse range of occupation numbers across the electron states and then relax towards quasi Fermi-Dirac distributions. The corresponding chemical potentials and the temperatures are determined by the amount and the energy of the electrons that have been excited. To visualize the statistics of the observed Fermi-Dirac distributions, \cref{fig:dataset}\,(c) and (d) plot histograms of the electron density (the thermodynamic conjugate to the chemical potential) and the electron temperature. We observe densities from $0.03\u{nm}^{-1}$ to $0.25\u{nm}^{-1}$ and temperatures from $250\u{K}$ to $430\u{K}$. Note that those electron temperatures only characterize quasi-equilibrium distributions, which have not thermalized with respect the phonon system yet.

\section{Singular Value Decomposition Based Dimensionality Reduction} \label{sec:SVD}

\begin{figure}[htbp]
\centering
\includegraphics[width=\linewidth]{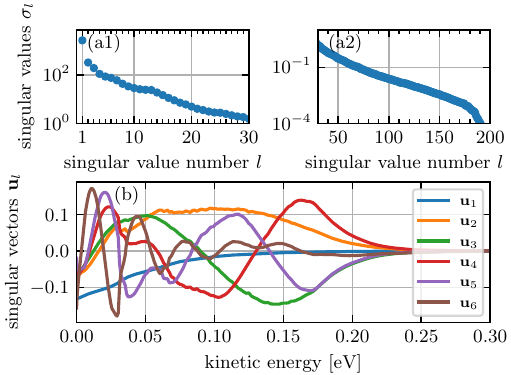}
\caption{Singular value spectrum split into (a1) and (a2), and the first six singular vectors (b) obtained from the SVD of the training data set with $d_s = 200$.
 }
\label{fig:SVD}
\end{figure}

In this section, we briefly illustrate how the singular value decomposition (SVD) based dimensionality reduction applies to our data. On that account, we use our standard data set with 1000 trajectories and $d_s=200$ (s. \cref{fig:dataset}), and compute the SVD according to \cref{eq:SVD}. 
The resulting singular value spectrum $\sigma_l$ is plotted in \cref{fig:SVD}\,(a1) and (a2) and the first six left singular vectors $\mathbf{u}_l$ are plotted in \cref{fig:SVD}\,(b).

The singular spectrum exhibits the typical s-shape of full-rank data matrices. The biggest jump between two adjacent singular values can be observed between the first and the second. The high energy found in the first singular value thus highlights the importance of the first mode (left singular vector) $\mathbf{u}_1$ for the expansion of the full electron state $f_k$. As presented in \cref{fig:SVD}\,(b) by the blue line, this mode resembles the electron quasi-equilibrium state, which is a good description of the later part of most trajectories (s. \cref{fig:dataset}\,(b)). The subsequent modes, however, to not intuitively resemble physical features of the system, but rather portray exponential decaying oscillations with increasing complexity, i.e. number of extrema/roots.

\begin{figure}[htbp]
\centering
\includegraphics[width=\linewidth]{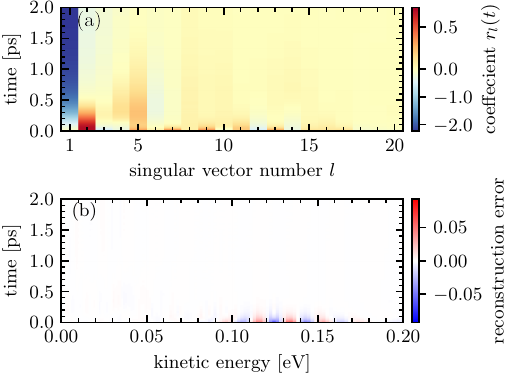}
\caption{Compressed representation of the dynamics shown in \cref{fig:dynamics} (a) and the corresponding reconstruction error (b).
Dimensionality reduction is achieved by expanding the electron state $f_k(t)$ in the first $r$ left singular vectors $\mathbf{u}_l$. The time evolution of the coefficients $r_l(t)$ is color-coded along the verticals of (a). The difference of the occupation numbers between the reconstruction and the ground truth produces the rms reconstruction error $\epsilon_\rm{rms} = 0.0024$ and the maximum error $\epsilon_\rm{max} = 0.0952$. Parameters: $d_s = 200$ and $d_r = 20$
 }
\label{fig:SVD_dyn_compression}
\end{figure}

To further demonstrate the dimensionality reduction, we project the example trajectory $f_k(t)$ presented in \cref{fig:dynamics} onto the first $r=20$ left singular vectors $\mathbf{u}_l$ to obtain the time evolution of the reduced state $\mathbf{r}(t)$. The evolution of the individual components $r_l(t)$ is color coded and presented along vertical lines in \cref{fig:SVD_dyn_compression}\,(a). The initial state (bottom row) is represented by a mixture of all modes, but the first, with similar weights. This indicates that the initial state is not well represented by a few of the chosen first 20 modes.
The relaxation towards the quasi-equilibrium distribution then causes the first coefficient to grow and all others to decay to values close to zero.

Lastly, the reconstruction error $f_k^\rm{recon}(t) - f_k(t)$ of the considered trajectory with $d_r=20$ is plotted in \cref{fig:SVD_dyn_compression}\,(b). Red and blue colors represent positive and negative deviations. Reconstruction errors can mostly be found in the early stage of the transient dynamics, where the distribution is close to its Gaussian initial conditions. There, the error alternates from positive to negative values, with frequencies given by the oscillatory structures of the left singular vectors $\mathbf{u}_l$. However, once the early stage is passed, the reconstruction errors become negligible as indicated by the vast white region.
Hence, further increasing the dimension of the reduced system state mostly benefits the accurate description of the early stage system states.

\section{Autoregression Error Statistics} \label{sec:error_statistics}

\begin{figure}[htbp]
\centering
\includegraphics[width=\linewidth]{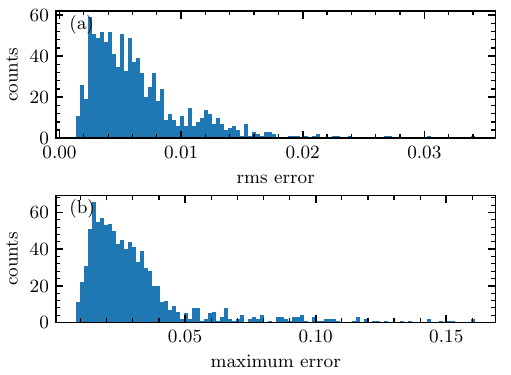}
\caption{Autoregression error statistics from a ten-fold cross validation procedure. Model with SVD based dimensionality reduction and polynomial features up to second order. Other parameters: $\ell=2$, $d_s = 200$, $d_r = 20$, $r_s = 0.1$, and $\alpha = 10^{-3.5}$
 }
\label{fig:pred_error_statistics}
\end{figure}

In this section, we briefly discuss the autoregression error statistics for the model with polynomial feature up to second order. We apply a ten-fold cross validation procedure (s. \cref{sec:benchmarking}) to our standard data set with $d_s=200$ (s. \cref{sec:training_data}) to obtain individual error scores for each of the 1000 trajectories. \Cref{fig:pred_error_statistics} shows a histogram of the rms error in (a) and a histogram of the maximum error in (b). The model hyper-parameters are given in the caption. Both error scores exhibit a similar qualitative behavior: Most errors cluster in a relatively small region and small fraction produces a long tail towards larger errors. Those outliers drive both the mean and the standard deviation to large values even though they are not representative of most of the errors. Nonetheless, we deliberately do not use outlier robust measures such as the median and the quartiles, because we want penalize outliers in the regression error scores, as they are used in the main body of the manuscript. In that sense, even a single test trajectory, for which the forecasting produces large errors, is highly undesirable for the considered applications in multi-physics simulations.

% \bibliography{export,references}

%merlin.mbs apsrev4-1.bst 2010-07-25 4.21a (PWD, AO, DPC) hacked
%Control: key (0)
%Control: author (72) initials jnrlst
%Control: editor formatted (1) identically to author
%Control: production of article title (-1) disabled
%Control: page (0) single
%Control: year (1) truncated
%Control: production of eprint (0) enabled
%

\end{document}